\documentclass[prb,twocolumn,showpacs,amssymb,superscriptaddress,longbibliography]{revtex4-2}
\usepackage{graphicx}
\usepackage{amsmath}
\usepackage{amsfonts}
\usepackage{amsthm}
\usepackage{amssymb}
\usepackage{amsbsy}
\usepackage{wasysym}
\usepackage{bm}
\usepackage{bbm}
\usepackage{mathrsfs}
\usepackage{color}
\usepackage{hyperref}
\usepackage{braket}
\usepackage{soul}
\setcitestyle{super}

\date{\today}
\begin{document}

\title{Hot band sound}

\author{Vir B. Bulchandani}
\affiliation{Department of
Physics and Astronomy, Rice University, 6100 Main Street
Houston, TX 77005, USA}

\author{David A. Huse}
\affiliation{Department of Physics, Princeton University, Princeton, NJ, 08544, USA}

\begin{abstract}
Chaotic lattice models at high temperature are generically expected to exhibit diffusive transport of all local conserved charges. Such diffusive transport is usually associated with overdamped relaxation of the associated currents. Here we show that by appropriately tuning the inter-particle interactions, lattice models of chaotic fermions at infinite temperature can be made to cross over from an overdamped regime of diffusion to an underdamped regime of ``hot band sound''. We study a family of one-dimensional spinless fermion chains with long-range density-density interactions, in which the damping time of sound waves can be made arbitrarily long even as an effective interaction strength is held fixed. Our results demonstrate that underdamped sound waves of charge density can arise within a single band, with strong interactions and far from integrability, and at very high temperature.
\end{abstract}

\maketitle
\section{Introduction} Understanding the transport properties of systems of strongly interacting fermions continues to pose a substantial challenge for theory. One prominent open problem is understanding the resistivity growth of so-called ``bad metals'' at temperatures above the Mott-Ioffe-Regel limit, where the mean free path is of order the lattice spacing, so a semiclassical description is no longer a good first approximation
\cite{SuperBadMetals,ResSatreview,MIRUniv}. While there is still no firm consensus on the origin of bad metal physics, strong electron-electron interactions are expected to play a significant role\cite{Tlinrev,Phillips}.

Identifying the specific consequences of strong electron-electron interactions in realistic condensed matter systems is often complicated by the presence of disorder and phonons. Recently, however, it has become possible to isolate purely ``electronic'' effects through cold-atom realizations of strongly interacting fermions in optical lattices, which have neither phonons nor disorder.  One such experiment explored bad metal physics in the two-dimensional single-band Fermi-Hubbard model, extending to temperatures that are large compared to the hopping strength\cite{Bakr}. An unexpected finding in this experiment was the observation of a transiently ballistic regime of underdamped ``band sound'' for low enough temperatures and short enough wavelengths. 

What do we mean by ``band sound''?  We mean underdamped density oscillations of interacting quantum lattice particles moving in a single band.  Motivated by the above-mentioned experimental results, this paper provides some theoretical justification for the phenomenon of such underdamped and thus transiently ballistic density modes in fermion lattice models.  In particular, we provide detailed analytical and numerical arguments that such underdamped band sound can occur in systems that are quantum chaotic and at very high (even infinite) temperature. 

Our results are surprising because the na{\"i}ve expectation for the hydrodynamics of conserved charges in quantum chaotic lattice systems at very high temperature is normal diffusion. This should be contrasted with continuum systems, for which momentum conservation generically gives rise to ballistic sound modes.  Similarly, quantum integrable systems on the lattice are well-known to exhibit ballistic and anomalous regimes of transport\cite{Zotos_1997,Znidaric,SuperdiffLattice,agrawal2020anomalous,scheie2021detection,Blochexp,Bulchandani_2021}, while chaotic lattice systems at low temperature can also exhibit anomalous transport by virtue of their proximity to integrable points\cite{Bulchandani_2021,BKM}. However, for systems that are on a lattice, far from integrability and at high temperature, no such behaviour is expected.

Our main result is the construction of a family of chaotic, interacting spinless fermion chains in which the decay rate of the charge current can be made arbitrarily small as the interaction range is increased, all the while keeping the magnitude of the interaction term fixed. We explain this rather unusual phenomenology through a quantum Boltzmann equation treatment. Although the latter is only strictly justifiable at weak coupling, it leads to a simple physical interpretation of the models we obtain: these models exhibit strong scattering that is concentrated at very low momentum transfer. This both yields an efficient pathway to chaos and implies that Umklapp scattering is suppressed, so that the single-particle phase-space distribution function itself behaves like a locally conserved, hydrodynamic degree of freedom for long times. The kinetic theory of this ``phase-space hydrodynamics'' qualitatively resembles kinetic theories of quasiparticle dynamics in quantum integrable systems\cite{CastroAlvaredo,Fagotti,BVKM1,DoyonSoliton,DephNoise} and the Landau kinetic equation that describes zero sound in cold Fermi liquids\cite{lifshitz2013statistical}. Thus a novel prediction of this work is that such phase-space hydrodynamics can emerge both far from integrability and at high temperature.

The paper is structured as follows. We first motivate our approach before introducing the charge-conserving lattice models that will be considered in this paper. We then formulate a variational problem for minimizing the decay of the charge current in this family of models, and present a class of exact solutions to this variational problem. Finally, we simulate the resulting ``optimal models'' numerically, and show that they exhibit signatures of both hot band sound and many-body quantum chaos.

\section{Motivation from the Fermi-Hubbard model}
Previous experimental work on the two-dimensional Fermi-Hubbard model~\cite{Bakr} found that the time-evolution of the local charge density $n(\mathbf{x},t)$ was well characterized by the pair of equations
\begin{align}
\label{eq:FHcons}
\partial_t n + \nabla \cdot \mathbf{j} &= 0,\\
\label{eq:FHdecay}
\partial_t \mathbf{j} &= - \Gamma(\mathbf{j}+D \nabla n)
\end{align}
with the charge diffusion constant $D$ and the current decay rate $\Gamma$ treated as fitting parameters. At large wavenumbers $k \gg \sqrt{\Gamma/4D}$, this system of equations describes ballistic propagation of an underdamped sound-like mode with propagation speed $c^2 = \Gamma D$. In the hydrodynamic regime of small wavenumbers $k \ll \sqrt{\Gamma/4D}$, this transient ballistic dynamics gives way to normal diffusion of charge.

Textbook hydrodynamic theory implicitly assumes that $\Gamma = \infty$; the charge-current density is not a local conserved charge of the Fermi-Hubbard Hamiltonian and should not give rise to a long-lived hydrodynamic mode. The experimental observation of a transiently ballistic charge mode at sufficiently short wavelengths, consistent with a finite current relaxation rate $\Gamma < \infty$, thus lies beyond the scope of conventional hydrodynamics and linear response theory, although some recent work has obtained estimates for $\Gamma$ and $D$ from kinetic theory~\cite{Mueller,Vucicevic}.

In this paper, we pursue a different route towards understanding how such ``quasihydrodynamic''~\cite{Quasihydro} behaviour of current densities can arise, which does not assume the phenomenological model Eq. \eqref{eq:FHdecay}. Specifically, we introduce a microscopic estimate of the relaxation rate of the charge current and construct lattice models in which this estimate can be made arbitrarily small. We verify the effectiveness of our proposal by simulating these models numerically and recovering underdamped charge relaxation, as seen in the Fermi-Hubbard model at short wavelengths~\cite{Bakr}, even at the very longest wavelengths accessible via exact diagonalization (see Fig. \ref{Fig1}). We focus specifically on infinite temperature and half-filling, where long-lived ballistic modes of charge are intuitively most surprising, but we expect by continuity that the models we obtain exhibit similar behaviour for large but finite temperatures; this expectation is borne out numerically\cite{SuppMat}.

\section{Models with arbitrarily long-lived charge currents}
For simplicity, we focus on translation-invariant, charge-conserving spinless fermion chains with density-density interactions
\begin{align}
\nonumber H = \hat{H}_0 + \hat{V}, \quad \hat{H}_0 = -\sum_{x,x'=1}^L t_{|x-x'|}\hat{c}_{x'}^\dagger \hat{c}_x,\\
\label{eq:modelspace}
\hat{V} =  \sum_{x,x'=1}^L U_{|x-x'|} (\hat{n}_{x'}-1/2)(\hat{n}_x-1/2)
\end{align}
where $t_0=U_0=0$, we assume periodic boundary conditions $x \equiv x+L$, and we set $t_{|x-x'|} = U_{|x-x'|} = 0$ for $|x-x'| \geq \lfloor L/2 \rfloor$ to avoid double counting. For site $x$, the onsite charge density $\hat{n}_x$ satisfies the operator-valued continuity equation $\partial_t \hat{n}_x + \hat{j}_{x+1} - \hat{j}_x = 0$, where the charge current operator $\hat{j}_x = i\sum_{r>0} \sum_{x'=x}^{x+r-1} t_r (\hat{c}_{x'}^\dagger \hat{c}_{x'-r} - \hat{c}_{x'-r}^\dagger \hat{c}_{x'})$ is independent of the interaction strength. Let us define total charges and currents $\hat{N} = \sum_{x} \hat{n}_x$ and $\hat{J} = \sum_{x} \hat{j}_x$. In general $\hat{N}$ is conserved but $\hat{J}$ is not, leading to diffusive transport of charge. Nevertheless, if $\hat{J}$ decays very slowly, the possibility arises of an underdamped and thus transiently ballistic ``sound mode'' of lattice fermions at high temperature and short enough wavelengths. 

We seek to realize this regime by tuning the form of the interactions. It will be useful to write the charge current explicitly in terms of Fourier modes $\hat{c}_x = \frac{1}{\sqrt{L}}\sum_k e^{ikx}\hat{c}_k$ as $
\hat{J} = \sum_k v_k \hat{c}_k^\dagger \hat{c}_k$, 
where the group velocity $v_k = \sum_{r>0} 2rt_r \sin{kr}$. This operator is manifestly conserved under the non-interacting dynamics due to $\hat{H}_0$, but decays due to the interaction term $\hat{V}$. In real space, we find that
\begin{align}
\nonumber &\dot{\hat{J}} = -\sum_{r>0}2rt_r\sum_x \\
&\sum_{y\neq x,x+r} \left(U_{|y-x|} - U_{|y-x+r|}\right)(\hat{n}_y-1/2) (\hat{c}^\dagger_x \hat{c}_{x-r} + \hat{c}_{x-r}^\dagger \hat{c}_x).
\end{align}
We would like to minimize the decay rate of $\hat{J}$ at infinite temperature. At long times, this decay rate is captured by the charge diffusion constant, which may be computed from the integral $\int_0^{\infty}dt\,\langle \hat{J}(t) \hat{J}(0)\rangle_{\beta=0}$ via the Kubo formula~\cite{mahan2013many}. However, the Kubo formula is not analytically tractable for generic chaotic systems, and the underdamped relaxation we seek to capture is moreover a short-time effect. It therefore suffices to focus on the short-time decay rate of $\hat{J}(t)$. Even at short times, there is no canonical microscopic definition of this decay rate, given that the expectation value of the rate of change of the current operator vanishes, $\langle \dot{\hat{J}}(t) \rangle_{\beta=0} = 0$, and $
\langle \hat{J} \hat{J}(t)\rangle_{\beta=0} =  \langle \hat{J} \hat{J}(-t)\rangle_{\beta=0}$ by cyclicity of the trace. The latter equation is a simple case of the Kubo-Martin-Schwinger relation, and here implies time-reversal symmetry of $\langle \hat{J} \hat{J}(t)\rangle_{\beta=0}$ and therefore vanishing of the correlation function $
\langle \hat{J} \dot{\hat{J}}(t)\rangle_{\beta=0} = 0$. 

Since these simple-minded estimates of the decay rate fail, we instead focus on the \emph{variance} of the rate of change of the current operator $\langle \dot{\hat{J}}(t)^2 \rangle_{\beta=0}$, which is generically non-zero. This leads to a simple and microscopic dimensionful estimate of the lifetime of the charge-current operator, given by
\begin{equation}
\label{eq:deftaueff}
\tau_{\mathrm{eff}}^{-1} = \sqrt{\frac{\langle \dot{\hat{J}}(t)^2 \rangle_{\beta=0}}{\langle \hat{J}(t)^2 \rangle_{\beta=0}}},
\end{equation}
which is independent of time. We now show that this physically reasonable definition of the decay rate is sufficient to predict hot band sound. This approach was used previously to construct slowly relaxing operators in spin chains\cite{Kim_2015}.

To maximally enhance band sound, we must minimize the decay rate $\tau_{\mathrm{eff}}^{-1}$ over the space of models Eq. \eqref{eq:modelspace}. To this end, we minimize the exact quantity
\begin{align}
\label{eq:exactvariance}
\langle \dot{\hat{J}}\dot{\hat{J}}\rangle_{\beta=0} = \frac{1}{2} \sum_{r>0} r^2 t_r^2 \sum_x \sum_{y\neq x,x-r}\big(U_{|y-x|}-U_{|y-x+r|}\big)^2.
\end{align}
To fix the denominator in Eq. \eqref{eq:deftaueff}, we set the variance of the current operator $\hat{J}$ equal to a characteristic scale $\sigma_J^2 = L/2$. To avoid trivial solutions with no interactions we similarly fix the variance of the interaction operator $\hat{V}$ to equal a characteristic fluctuation scale $\sigma_V^2 = L/4$. (Note that both scales are set by unit nearest-neighbour hoppings and interactions $t_1=U_1=1$, and that both $\hat{V}$ and $\hat{J}$ are traceless.) This yields the Lagrangian
\begin{align}
\nonumber
&\mathcal{L}(t_r,U_r,\lambda_1,\lambda_2) = \langle \dot{\hat{J}}\dot{\hat{J}}\rangle_{\beta=0} \\
\label{eq:Lfn}
+&\lambda_1  \left(\langle \hat{V}^2\rangle_{\beta=0} - \sigma_V^2\right) 
+\lambda_2 \left(\langle \hat{J}^2\rangle_{\beta=0} -\sigma_J^2\right),
\end{align}
which is a function of the hopping and interaction strengths at each range $r$. Restricting to nearest-neighbour hopping $t_1=1$, $t_r = 0$ for $r>1$, this optimization problem becomes exactly solvable and for each interaction range $R \geq 1$, we find that the density-density interactions minimizing $\tau_{\mathrm{eff}}^{-1}$ take the form~\cite{SuppMat}
\begin{equation}
\label{eq:trueoptimum}
U^*_n(R) = \frac{2}{\sqrt{2R+1}} \cos{\frac{\pi(n-1/2)}{(2R+1)}}, \quad 1 \leq n \leq R,
\end{equation}
with an associated decay rate $
\tau_{\mathrm{eff}}^{-1}  = 2 \sqrt{1 -\cos \left(\frac{\pi}{2R+1}\right)}$
that is independent of the system size, since both the numerator and denominator in Eq. \eqref{eq:deftaueff} are extensive. For $R=1$, this is an integrable spinless fermion model equivalent to the spin-$1/2$ Heisenberg chain\footnote{This model in fact exhibits superdiffusive charge transport\cite{Znidaric,SuperdiffLattice,scheie2021detection,Blochexp,Bulchandani_2021}. Tuning the effective XXZ anisotropy $\Delta=U_1$ of the integrable reference point in our constrained optimization problem is equivalent to tuning the value of the interaction constraint $\sigma_V^2$. We have checked that even when $\sigma_V^2/L<1/4$, which corresponds to an anisotropy $|\Delta| <1$ and hence ballistic charge transport for the integrable reference point, the optimal models with $R>1$ continue to exhibit a clear short-time enhancement of band sound compared to the integrable reference point with $R=1$. One interesting corollary of our analysis with $\Delta=1$ is that spin superdiffusion in the spin-$1/2$ Heisenberg chain occurs \emph{despite} the presence of substantial Umklapp scattering.}. For $R>1$ we obtain a family of models (henceforth ``optimal models'') for which the rate of current decay decreases with $R$. In the regime of interactions at a large but finite range, $R \gg 1$, the asymptotic decay rate is given by
\begin{equation}
\label{eq:teffasymp}
\tau_{\mathrm{eff}}^{-1} \sim \frac{\pi}{\sqrt{2}}\frac{1}{R} \to 0, \quad R \to \infty.
\end{equation}
Thus the decay of the charge current in this family of models can be made arbitrarily slow as the interaction range $R$ is increased. 
This expression should be seen as upper bound on the true relaxation rate. Nevertheless, a more detailed (albeit approximate) quantum Boltzmann equation treatment\cite{SuppMat} predicts that the fastest relaxation in this system arises from local diffusion in pseudomomentum space, with the same scaling $\tau^{-1} \propto 1/R$ as Eq. \eqref{eq:teffasymp}.

\section{Numerical results} 
\begin{figure}[t]
    \centering
    \includegraphics[width=0.99\linewidth]{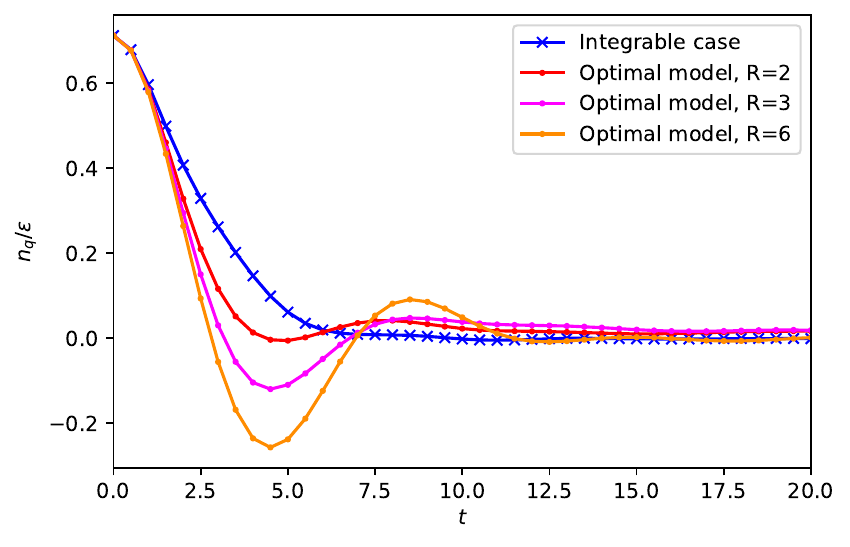}
    \caption{Decay of an initial density modulation in a system of $L=14$ sites at half-filling, near infinite temperature and for various optimal models with $1 \leq R < \lfloor L/2 \rfloor$. We include the integrable model with $R=1$ for comparison. It is clear that the level of damping decreases as $R$ increases.}
    \label{Fig1}
\end{figure}
To verify that the optimal models obtained above indeed exhibit underdamped sound modes, we first examine the relaxation of weak density modulations in real space. Therefore consider the initial condition
\begin{equation}
\label{eq:initialdensmod}
\hat{\rho}(0) = \frac{1}{Z}\left(1 + \epsilon \sum_{x=1}^{L} \sin{(qx)} (\hat{n}_x - \langle \hat{n}_x \rangle_{\beta=0})\right),
\end{equation}
with $\epsilon = 0.01$, $q = 2\pi /L$ and $Z=\mathrm{tr}[\hat{\rho}(0)]$, let it evolve numerically under Schr{\"o}dinger evolution $\hat{\rho}(t) = e^{-i\hat{H} t} \hat{\rho}(0) e^{i\hat{H} t}$ and look at the time evolution of the lowest Fourier mode of the charge density, namely $n_q(t)  = \sqrt{(2/L)}\sum_{x=1}^L \sin{(q x)} \mathrm{tr}[\hat{\rho}(t)\hat{n}_x]$. The resulting dynamics for optimal models with interaction ranges $R \in \{1,2,3,6\}$ is shown in Fig. \ref{Fig1} for half-filled chains on $L=14$ sites. There is a clear decrease in the level of damping as one moves to larger interaction ranges $R$, as expected.

\begin{figure}[t]
    \centering
    \includegraphics[width=0.99\linewidth]{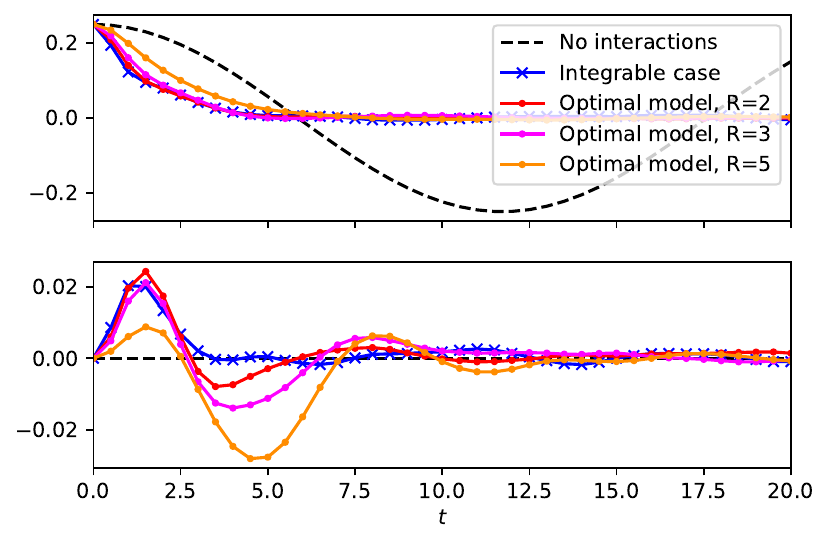}
    \caption{Decay and growth of pseudomomentum correlations
    $\mathrm{Re}[\langle \hat{c}_{k=\frac{2\pi}{L}}^\dagger \hat{c}_{0} \rangle_{\hat{\rho}(t)}]/\epsilon$ (\textit{top}) and $\mathrm{Re}[\langle \hat{c}_{k=\frac{6 \pi}{L}}^\dagger \hat{c}_{k=\frac{4\pi}{L}} \rangle_{\hat{\rho}(t)}]/\epsilon$ (\textit{bottom}) in the time evolution of the initial state Eq. \eqref{eq:cohinit}, for an $L=12$ site system. The presence of interactions in the optimal models leads to a coherent transfer of correlations from small to large pseudomomenta whose magnitude increases with $R$. This behaviour differs markedly from the non-interacting case.}
    \label{Fig2}
\end{figure}
We next present numerical evidence that underdamped charge relaxation in these models is associated with interactions, rather than proximity to a non-interacting system as Eq. \eqref{eq:teffasymp} might suggest. To probe this physics, we consider optimal models on $L=12$ sites, initialized in a state that has weak coherence between pseudomomentum modes with $k = \pm 2\pi /L$ and $k=0$, but no other correlations between distinct pseudomomenta, namely
\begin{equation}
\label{eq:cohinit}
\hat{\rho}(0) = \frac{1}{2^L}\left(1 + \epsilon(\hat{c}_{k=0}^\dagger (\hat{c}_{k=\frac{2\pi}{L}} + \hat{c}_{k=-\frac{2\pi}{L}}) + \mathrm{h.c.})\right)
\end{equation}
where we again set the small parameter $\epsilon = 0.01$. In the absence of interactions, the time-evolution of this density matrix is given simply by $
\hat{\rho}(t) = \frac{1}{2^L}\left(1 + \epsilon(e^{i\omega t}\hat{c}_{k=0}^\dagger( \hat{c}_{k=\frac{2\pi}{L}} + \hat{c}_{k=-\frac{2\pi}{L}}) + \mathrm{h.c.})\right)$, where $\omega = 2 (1-\cos{2\pi/L}) \sim 1/L^2$ for large $L$. Thus the expectation value $
\langle \hat{c}_{k=\frac{2\pi}{L}}^\dagger \hat{c}_{k=0} \rangle_{\hat{\rho}(t)} = \frac{\epsilon}{4} e^{i\omega t}$
while for higher-pseudomomentum correlations, e.g. for $Lk/2\pi=2,3$ we have $\langle \hat{c}_{k=\frac{6 \pi}{L}}^\dagger \hat{c}_{k=\frac{4\pi}{L}} \rangle_{\hat{\rho}(t)} = 0$.

In the presence of interactions, scattering between pseudomomenta should deplete correlations between the lowest lying pseudomomentum modes and enhance correlations between higher pseudomomentum modes. A coherent transfer of such correlations, indicative of normal scattering rather than Umklapp scattering, is demonstrated numerically in Fig. \ref{Fig2}.  Note that the resulting underdamped sound mode in the interacting system has a substantially higher frequency than the oscillations of the non-interacting system produced by this initial state.

\begin{figure}[t]
    \centering
    \includegraphics[width=0.8\linewidth]{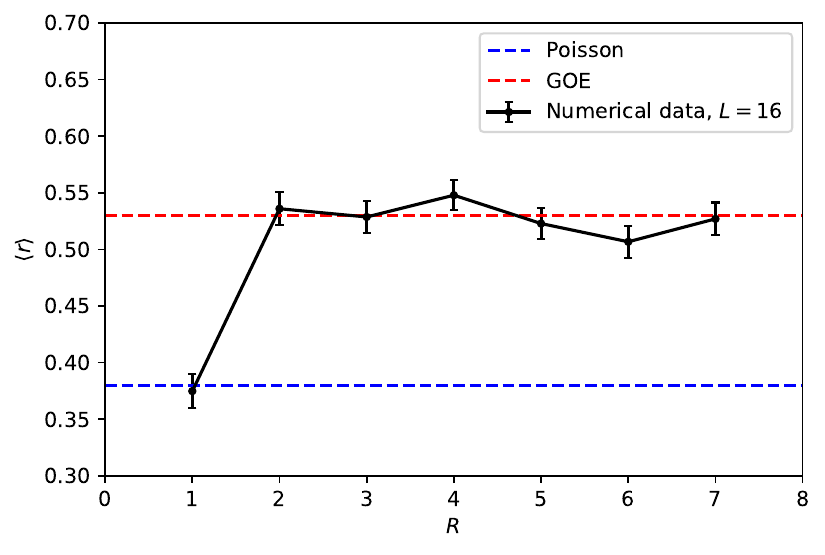}
    \caption{Numerical evaluation of the $\langle r \rangle$ statistic for various optimal models, near half-filling with $L=16$ sites and $M=7$ fermions. After projecting onto the sector with zero pseudomomentum and even parity to eliminate any remaining degeneracies, 340 states remain. Error bars denote standard error of the sample mean in computing $\langle r \rangle$. The numerical evidence clearly indicates that the optimal models are chaotic for $R>1$. Note that there is no apparent tendency towards integrability with increasing $R$.}
    \label{Fig3}
\end{figure}

We next turn to the question of whether these optimal models are quantum chaotic. To this end, we compute the $\langle r \rangle$ statistic\cite{rstat} for the optimal models at a relatively large numerically accessible system size, $L=16$.  The results are plotted in Fig. \ref{Fig3} and strongly suggest that the optimal models are chaotic for $R>1$, since the $\langle r \rangle$ statistic jumps from its expected Poisson value $\langle r \rangle \approx 0.38$ at $R=1$, indicating quantum integrability, to its Gaussian Orthogonal Ensemble (GOE) value $\langle r \rangle \approx 0.53$ for $R > 1$, indicating quantum chaos.

\section{Conclusion} 
We have constructed a family of fermionic lattice models that exhibit both infinite-temperature band sound and signatures of many-body quantum chaos at numerically accessible system sizes. We have further shown that the decay rate of the charge current in these models can be made arbitrarily small by increasing the range of the density-density interaction, while keeping the interactions strong. We expect similar conclusions to hold for the corresponding models one could make in higher spatial dimensions $d>1$.

Our study was motivated by the experimental observation of a transient ballistic mode of charge density in a cold atom, single-band Fermi-Hubbard model\cite{Bakr}. One outstanding question is how the physics of hot band sound connects to these experimental observations for the 2D Fermi-Hubbard model. We believe that kinematic constraints on two-particle scattering are ultimately responsible for both phenomena and that these constraints are enhanced for the models studied above, which is why they exhibit band sound at longer wavelengths and much higher temperatures than the Fermi-Hubbard model. Indeed, the quantum Boltzmann equation for spinless fermion chains Eq. \eqref{eq:modelspace} with nearest-neighbor hopping predicts infinitely many long-lived slow modes~\cite{SuppMat} at all temperatures, analogous to the phenomenon of ``tomographic dynamics'' in 2D Fermi liquids at low temperatures~\cite{Ledwith_2019}. However, we defer a detailed discussion of the microscopic mechanism giving rise to band sound in the 2D Fermi-Hubbard model to future work.

The spinless fermion chains constructed in this paper are equivalent by a Jordan-Wigner transformation to spin-$1/2$ chains with nearest-neighbour exchange and long-range $ZZ$ interactions, in which band sound appears as long-lived ballistic modes of the local magnetization $\langle \hat{S}^z_i\rangle$ at high temperature. Systems of trapped ions naturally realize such long-range interacting spin-$1/2$ chains\cite{trapped_ion_review}. While constructing the specific interaction graph corresponding to an optimal model requires a high degree of tunability, such fine control is possible in principle\cite{Tunable,PhysRevX.13.041052}, and the qualitative effects we find should not require such precise optimality. This might provide one near-term experimental realization of the physics in this paper. An alternative approach, which has recently proved effective for probing the infinite-temperature dynamics of the spin-$1/2$ Heisenberg chain~\cite{rosenberg2024dynamics} and may be feasible for optimal models with small interaction ranges $R$, would be to discretize the spin-chain time evolution on a digital quantum computer.

We note that similar ``quasiballistic'' behaviour was previously observed numerically in spin-$1/2$ chains with both long-range exchange and long-range interactions\cite{Quasiballistic,PhysRevB.110.014308}, although the latter models have no fermionic counterpart and therefore do not exhibit band sound in the sense of this paper. Nevertheless, some essential features of such quasiballistic spin transport can be explained by adapting the arguments above~\cite{khpy-d9s7}. Meanwhile, transiently ballistic energy transport has recently been observed in relatively generic quantum lattice models~\cite{Ljubotina_2023,PhysRevB.110.134308,mark2025observationballisticplasmamemory}. It would be interesting to understand how far either of these distinct spin and energy transport phenomena can be related to the low-momentum-transfer scattering of quasiparticles that gives rise to hot band sound.

\section{Acknowledgments} 
We thank Fabian Essler, Curt von Keyserlingk, Sarang Gopalakrishnan and Hyunsoo Ha for helpful discussions. V.B.B. was supported by a Fellowship at the Princeton Center for Theoretical Science during part of this work. D.A.H. was supported in part by NSF QLCI grant OMA-2120757. The simulations reported in this paper were performed on the Della cluster at Princeton University and on the NOTS cluster at Rice University.
\bibliography{bibl.bib}

\begin{thebibliography}{43}%
\makeatletter
\providecommand \@ifxundefined [1]{%
 \@ifx{#1\undefined}
}%
\providecommand \@ifnum [1]{%
 \ifnum #1\expandafter \@firstoftwo
 \else \expandafter \@secondoftwo
 \fi
}%
\providecommand \@ifx [1]{%
 \ifx #1\expandafter \@firstoftwo
 \else \expandafter \@secondoftwo
 \fi
}%
\providecommand \natexlab [1]{#1}%
\providecommand \enquote  [1]{``#1''}%
\providecommand \bibnamefont  [1]{#1}%
\providecommand \bibfnamefont [1]{#1}%
\providecommand \citenamefont [1]{#1}%
\providecommand \href@noop [0]{\@secondoftwo}%
\providecommand \href [0]{\begingroup \@sanitize@url \@href}%
\providecommand \@href[1]{\@@startlink{#1}\@@href}%
\providecommand \@@href[1]{\endgroup#1\@@endlink}%
\providecommand \@sanitize@url [0]{\catcode `\\12\catcode `\$12\catcode
  `\&12\catcode `\#12\catcode `\^12\catcode `\_12\catcode `\%12\relax}%
\providecommand \@@startlink[1]{}%
\providecommand \@@endlink[0]{}%
\providecommand \url  [0]{\begingroup\@sanitize@url \@url }%
\providecommand \@url [1]{\endgroup\@href {#1}{\urlprefix }}%
\providecommand \urlprefix  [0]{URL }%
\providecommand \Eprint [0]{\href }%
\providecommand \doibase [0]{https://doi.org/}%
\providecommand \selectlanguage [0]{\@gobble}%
\providecommand \bibinfo  [0]{\@secondoftwo}%
\providecommand \bibfield  [0]{\@secondoftwo}%
\providecommand \translation [1]{[#1]}%
\providecommand \BibitemOpen [0]{}%
\providecommand \bibitemStop [0]{}%
\providecommand \bibitemNoStop [0]{.\EOS\space}%
\providecommand \EOS [0]{\spacefactor3000\relax}%
\providecommand \BibitemShut  [1]{\csname bibitem#1\endcsname}%
\let\auto@bib@innerbib\@empty
\bibitem [{\citenamefont {Emery}\ and\ \citenamefont
  {Kivelson}(1995)}]{SuperBadMetals}%
  \BibitemOpen
  \bibfield  {author} {\bibinfo {author} {\bibfnamefont {V.~J.}\ \bibnamefont
  {Emery}}\ and\ \bibinfo {author} {\bibfnamefont {S.~A.}\ \bibnamefont
  {Kivelson}},\ }\bibfield  {title} {\bibinfo {title} {Superconductivity in bad
  metals},\ }\href {https://doi.org/10.1103/PhysRevLett.74.3253} {\bibfield
  {journal} {\bibinfo  {journal} {Phys. Rev. Lett.}\ }\textbf {\bibinfo
  {volume} {74}},\ \bibinfo {pages} {3253} (\bibinfo {year}
  {1995})}\BibitemShut {NoStop}%
\bibitem [{\citenamefont {Gunnarsson}\ \emph {et~al.}(2003)\citenamefont
  {Gunnarsson}, \citenamefont {Calandra},\ and\ \citenamefont
  {Han}}]{ResSatreview}%
  \BibitemOpen
  \bibfield  {author} {\bibinfo {author} {\bibfnamefont {O.}~\bibnamefont
  {Gunnarsson}}, \bibinfo {author} {\bibfnamefont {M.}~\bibnamefont
  {Calandra}},\ and\ \bibinfo {author} {\bibfnamefont {J.~E.}\ \bibnamefont
  {Han}},\ }\bibfield  {title} {\bibinfo {title} {Colloquium: Saturation of
  electrical resistivity},\ }\href {https://doi.org/10.1103/RevModPhys.75.1085}
  {\bibfield  {journal} {\bibinfo  {journal} {Rev. Mod. Phys.}\ }\textbf
  {\bibinfo {volume} {75}},\ \bibinfo {pages} {1085} (\bibinfo {year}
  {2003})}\BibitemShut {NoStop}%
\bibitem [{\citenamefont {Hussey}\ \emph {et~al.}(2004)\citenamefont {Hussey},
  \citenamefont {Takenaka},\ and\ \citenamefont {Takagi}}]{MIRUniv}%
  \BibitemOpen
  \bibfield  {author} {\bibinfo {author} {\bibfnamefont {N.~E.}\ \bibnamefont
  {Hussey}}, \bibinfo {author} {\bibfnamefont {K.}~\bibnamefont {Takenaka}},\
  and\ \bibinfo {author} {\bibfnamefont {H.}~\bibnamefont {Takagi}},\
  }\bibfield  {title} {\bibinfo {title} {{Universality of the
  Mott–Ioffe–Regel limit in metals}},\ }\href
  {https://doi.org/10.1080/14786430410001716944} {\bibfield  {journal}
  {\bibinfo  {journal} {Philosophical Magazine}\ }\textbf {\bibinfo {volume}
  {84}},\ \bibinfo {pages} {2847} (\bibinfo {year} {2004})},\ \Eprint
  {https://arxiv.org/abs/https://doi.org/10.1080/14786430410001716944}
  {https://doi.org/10.1080/14786430410001716944} \BibitemShut {NoStop}%
\bibitem [{\citenamefont {Hartnoll}\ and\ \citenamefont
  {Mackenzie}(2021)}]{Tlinrev}%
  \BibitemOpen
  \bibfield  {author} {\bibinfo {author} {\bibfnamefont {S.~A.}\ \bibnamefont
  {Hartnoll}}\ and\ \bibinfo {author} {\bibfnamefont {A.~P.}\ \bibnamefont
  {Mackenzie}},\ }\href {https://doi.org/10.48550/ARXIV.2107.07802} {\bibinfo
  {title} {Planckian dissipation in metals}} (\bibinfo {year} {2021}),\ \Eprint
  {https://arxiv.org/abs/2107.07802} {arXiv:2107.07802} \BibitemShut {NoStop}%
\bibitem [{\citenamefont {Phillips}\ \emph {et~al.}(2022)\citenamefont
  {Phillips}, \citenamefont {Hussey},\ and\ \citenamefont
  {Abbamonte}}]{Phillips}%
  \BibitemOpen
  \bibfield  {author} {\bibinfo {author} {\bibfnamefont {P.~W.}\ \bibnamefont
  {Phillips}}, \bibinfo {author} {\bibfnamefont {N.~E.}\ \bibnamefont
  {Hussey}},\ and\ \bibinfo {author} {\bibfnamefont {P.}~\bibnamefont
  {Abbamonte}},\ }\bibfield  {title} {\bibinfo {title} {Stranger than metals},\
  }\href {https://doi.org/10.1126/science.abh4273} {\bibfield  {journal}
  {\bibinfo  {journal} {Science}\ }\textbf {\bibinfo {volume} {377}},\ \bibinfo
  {pages} {eabh4273} (\bibinfo {year} {2022})},\ \Eprint
  {https://arxiv.org/abs/https://www.science.org/doi/pdf/10.1126/science.abh4273}
  {https://www.science.org/doi/pdf/10.1126/science.abh4273} \BibitemShut
  {NoStop}%
\bibitem [{\citenamefont {Brown}\ \emph {et~al.}(2019)\citenamefont {Brown},
  \citenamefont {Mitra}, \citenamefont {Guardado-Sanchez}, \citenamefont
  {Nourafkan}, \citenamefont {Reymbaut}, \citenamefont {Hébert}, \citenamefont
  {Bergeron}, \citenamefont {Tremblay}, \citenamefont {Kokalj}, \citenamefont
  {Huse}, \citenamefont {Schauß},\ and\ \citenamefont {Bakr}}]{Bakr}%
  \BibitemOpen
  \bibfield  {author} {\bibinfo {author} {\bibfnamefont {P.~T.}\ \bibnamefont
  {Brown}}, \bibinfo {author} {\bibfnamefont {D.}~\bibnamefont {Mitra}},
  \bibinfo {author} {\bibfnamefont {E.}~\bibnamefont {Guardado-Sanchez}},
  \bibinfo {author} {\bibfnamefont {R.}~\bibnamefont {Nourafkan}}, \bibinfo
  {author} {\bibfnamefont {A.}~\bibnamefont {Reymbaut}}, \bibinfo {author}
  {\bibfnamefont {C.-D.}\ \bibnamefont {Hébert}}, \bibinfo {author}
  {\bibfnamefont {S.}~\bibnamefont {Bergeron}}, \bibinfo {author}
  {\bibfnamefont {A.-M.~S.}\ \bibnamefont {Tremblay}}, \bibinfo {author}
  {\bibfnamefont {J.}~\bibnamefont {Kokalj}}, \bibinfo {author} {\bibfnamefont
  {D.~A.}\ \bibnamefont {Huse}}, \bibinfo {author} {\bibfnamefont
  {P.}~\bibnamefont {Schauß}},\ and\ \bibinfo {author} {\bibfnamefont {W.~S.}\
  \bibnamefont {Bakr}},\ }\bibfield  {title} {\bibinfo {title} {{Bad metallic
  transport in a cold atom Fermi-Hubbard system}},\ }\href
  {https://doi.org/10.1126/science.aat4134} {\bibfield  {journal} {\bibinfo
  {journal} {Science}\ }\textbf {\bibinfo {volume} {363}},\ \bibinfo {pages}
  {379} (\bibinfo {year} {2019})},\ \Eprint
  {https://arxiv.org/abs/https://www.science.org/doi/pdf/10.1126/science.aat4134}
  {https://www.science.org/doi/pdf/10.1126/science.aat4134} \BibitemShut
  {NoStop}%
\bibitem [{\citenamefont {Zotos}\ \emph {et~al.}(1997)\citenamefont {Zotos},
  \citenamefont {Naef},\ and\ \citenamefont {Prelovsek}}]{Zotos_1997}%
  \BibitemOpen
  \bibfield  {author} {\bibinfo {author} {\bibfnamefont {X.}~\bibnamefont
  {Zotos}}, \bibinfo {author} {\bibfnamefont {F.}~\bibnamefont {Naef}},\ and\
  \bibinfo {author} {\bibfnamefont {P.}~\bibnamefont {Prelovsek}},\ }\bibfield
  {title} {\bibinfo {title} {Transport and conservation laws},\ }\href
  {https://doi.org/10.1103/physrevb.55.11029} {\bibfield  {journal} {\bibinfo
  {journal} {Physical Review B}\ }\textbf {\bibinfo {volume} {55}},\ \bibinfo
  {pages} {11029} (\bibinfo {year} {1997})}\BibitemShut {NoStop}%
\bibitem [{\citenamefont {\ifmmode \check{Z}\else
  \v{Z}\fi{}nidari\ifmmode~\check{c}\else \v{c}\fi{}}(2011)}]{Znidaric}%
  \BibitemOpen
  \bibfield  {author} {\bibinfo {author} {\bibfnamefont {M.}~\bibnamefont
  {\ifmmode \check{Z}\else \v{Z}\fi{}nidari\ifmmode~\check{c}\else
  \v{c}\fi{}}},\ }\bibfield  {title} {\bibinfo {title} {Spin transport in a
  one-dimensional anisotropic heisenberg model},\ }\href
  {https://doi.org/10.1103/PhysRevLett.106.220601} {\bibfield  {journal}
  {\bibinfo  {journal} {Phys. Rev. Lett.}\ }\textbf {\bibinfo {volume} {106}},\
  \bibinfo {pages} {220601} (\bibinfo {year} {2011})}\BibitemShut {NoStop}%
\bibitem [{\citenamefont {Ilievski}\ \emph {et~al.}(2018)\citenamefont
  {Ilievski}, \citenamefont {De~Nardis}, \citenamefont {Medenjak},\ and\
  \citenamefont {Prosen}}]{SuperdiffLattice}%
  \BibitemOpen
  \bibfield  {author} {\bibinfo {author} {\bibfnamefont {E.}~\bibnamefont
  {Ilievski}}, \bibinfo {author} {\bibfnamefont {J.}~\bibnamefont {De~Nardis}},
  \bibinfo {author} {\bibfnamefont {M.}~\bibnamefont {Medenjak}},\ and\
  \bibinfo {author} {\bibfnamefont {T.}~\bibnamefont {Prosen}},\ }\bibfield
  {title} {\bibinfo {title} {Superdiffusion in one-dimensional quantum lattice
  models},\ }\href {https://doi.org/10.1103/PhysRevLett.121.230602} {\bibfield
  {journal} {\bibinfo  {journal} {Phys. Rev. Lett.}\ }\textbf {\bibinfo
  {volume} {121}},\ \bibinfo {pages} {230602} (\bibinfo {year}
  {2018})}\BibitemShut {NoStop}%
\bibitem [{\citenamefont {Agrawal}\ \emph {et~al.}(2020)\citenamefont
  {Agrawal}, \citenamefont {Gopalakrishnan}, \citenamefont {Vasseur},\ and\
  \citenamefont {Ware}}]{agrawal2020anomalous}%
  \BibitemOpen
  \bibfield  {author} {\bibinfo {author} {\bibfnamefont {U.}~\bibnamefont
  {Agrawal}}, \bibinfo {author} {\bibfnamefont {S.}~\bibnamefont
  {Gopalakrishnan}}, \bibinfo {author} {\bibfnamefont {R.}~\bibnamefont
  {Vasseur}},\ and\ \bibinfo {author} {\bibfnamefont {B.}~\bibnamefont
  {Ware}},\ }\bibfield  {title} {\bibinfo {title} {{Anomalous low-frequency
  conductivity in easy-plane XXZ spin chains}},\ }\href@noop {} {\bibfield
  {journal} {\bibinfo  {journal} {Physical Review B}\ }\textbf {\bibinfo
  {volume} {101}},\ \bibinfo {pages} {224415} (\bibinfo {year}
  {2020})}\BibitemShut {NoStop}%
\bibitem [{\citenamefont {Scheie}\ \emph {et~al.}(2021)\citenamefont {Scheie},
  \citenamefont {Sherman}, \citenamefont {Dupont}, \citenamefont {Nagler},
  \citenamefont {Stone}, \citenamefont {Granroth}, \citenamefont {Moore},\ and\
  \citenamefont {Tennant}}]{scheie2021detection}%
  \BibitemOpen
  \bibfield  {author} {\bibinfo {author} {\bibfnamefont {A.}~\bibnamefont
  {Scheie}}, \bibinfo {author} {\bibfnamefont {N.}~\bibnamefont {Sherman}},
  \bibinfo {author} {\bibfnamefont {M.}~\bibnamefont {Dupont}}, \bibinfo
  {author} {\bibfnamefont {S.}~\bibnamefont {Nagler}}, \bibinfo {author}
  {\bibfnamefont {M.}~\bibnamefont {Stone}}, \bibinfo {author} {\bibfnamefont
  {G.}~\bibnamefont {Granroth}}, \bibinfo {author} {\bibfnamefont
  {J.}~\bibnamefont {Moore}},\ and\ \bibinfo {author} {\bibfnamefont
  {D.}~\bibnamefont {Tennant}},\ }\bibfield  {title} {\bibinfo {title}
  {{Detection of Kardar--Parisi--Zhang hydrodynamics in a quantum Heisenberg
  spin-1/2 chain}},\ }\href@noop {} {\bibfield  {journal} {\bibinfo  {journal}
  {Nature Physics}\ }\textbf {\bibinfo {volume} {17}},\ \bibinfo {pages} {726}
  (\bibinfo {year} {2021})}\BibitemShut {NoStop}%
\bibitem [{\citenamefont {Wei}\ \emph {et~al.}(2022)\citenamefont {Wei},
  \citenamefont {Rubio-Abadal}, \citenamefont {Ye}, \citenamefont {Machado},
  \citenamefont {Kemp}, \citenamefont {Srakaew}, \citenamefont {Hollerith},
  \citenamefont {Rui}, \citenamefont {Gopalakrishnan}, \citenamefont {Yao},
  \citenamefont {Bloch},\ and\ \citenamefont {Zeiher}}]{Blochexp}%
  \BibitemOpen
  \bibfield  {author} {\bibinfo {author} {\bibfnamefont {D.}~\bibnamefont
  {Wei}}, \bibinfo {author} {\bibfnamefont {A.}~\bibnamefont {Rubio-Abadal}},
  \bibinfo {author} {\bibfnamefont {B.}~\bibnamefont {Ye}}, \bibinfo {author}
  {\bibfnamefont {F.}~\bibnamefont {Machado}}, \bibinfo {author} {\bibfnamefont
  {J.}~\bibnamefont {Kemp}}, \bibinfo {author} {\bibfnamefont {K.}~\bibnamefont
  {Srakaew}}, \bibinfo {author} {\bibfnamefont {S.}~\bibnamefont {Hollerith}},
  \bibinfo {author} {\bibfnamefont {J.}~\bibnamefont {Rui}}, \bibinfo {author}
  {\bibfnamefont {S.}~\bibnamefont {Gopalakrishnan}}, \bibinfo {author}
  {\bibfnamefont {N.~Y.}\ \bibnamefont {Yao}}, \bibinfo {author} {\bibfnamefont
  {I.}~\bibnamefont {Bloch}},\ and\ \bibinfo {author} {\bibfnamefont
  {J.}~\bibnamefont {Zeiher}},\ }\bibfield  {title} {\bibinfo {title} {{Quantum
  gas microscopy of Kardar-Parisi-Zhang superdiffusion}},\ }\href
  {https://doi.org/10.1126/science.abk2397} {\bibfield  {journal} {\bibinfo
  {journal} {Science}\ }\textbf {\bibinfo {volume} {376}},\ \bibinfo {pages}
  {716} (\bibinfo {year} {2022})},\ \Eprint
  {https://arxiv.org/abs/https://www.science.org/doi/pdf/10.1126/science.abk2397}
  {https://www.science.org/doi/pdf/10.1126/science.abk2397} \BibitemShut
  {NoStop}%
\bibitem [{\citenamefont {Bulchandani}\ \emph {et~al.}(2021)\citenamefont
  {Bulchandani}, \citenamefont {Gopalakrishnan},\ and\ \citenamefont
  {Ilievski}}]{Bulchandani_2021}%
  \BibitemOpen
  \bibfield  {author} {\bibinfo {author} {\bibfnamefont {V.~B.}\ \bibnamefont
  {Bulchandani}}, \bibinfo {author} {\bibfnamefont {S.}~\bibnamefont
  {Gopalakrishnan}},\ and\ \bibinfo {author} {\bibfnamefont {E.}~\bibnamefont
  {Ilievski}},\ }\bibfield  {title} {\bibinfo {title} {Superdiffusion in spin
  chains},\ }\href {https://doi.org/10.1088/1742-5468/ac12c7} {\bibfield
  {journal} {\bibinfo  {journal} {Journal of Statistical Mechanics: Theory and
  Experiment}\ }\textbf {\bibinfo {volume} {2021}},\ \bibinfo {pages} {084001}
  (\bibinfo {year} {2021})}\BibitemShut {NoStop}%
\bibitem [{\citenamefont {Bulchandani}\ \emph {et~al.}(2020)\citenamefont
  {Bulchandani}, \citenamefont {Karrasch},\ and\ \citenamefont {Moore}}]{BKM}%
  \BibitemOpen
  \bibfield  {author} {\bibinfo {author} {\bibfnamefont {V.~B.}\ \bibnamefont
  {Bulchandani}}, \bibinfo {author} {\bibfnamefont {C.}~\bibnamefont
  {Karrasch}},\ and\ \bibinfo {author} {\bibfnamefont {J.~E.}\ \bibnamefont
  {Moore}},\ }\bibfield  {title} {\bibinfo {title} {Superdiffusive transport of
  energy in one-dimensional metals},\ }\href
  {https://doi.org/10.1073/pnas.1916213117} {\bibfield  {journal} {\bibinfo
  {journal} {Proceedings of the National Academy of Sciences}\ }\textbf
  {\bibinfo {volume} {117}},\ \bibinfo {pages} {12713} (\bibinfo {year}
  {2020})}\BibitemShut {NoStop}%
\bibitem [{\citenamefont {Castro-Alvaredo}\ \emph {et~al.}(2016)\citenamefont
  {Castro-Alvaredo}, \citenamefont {Doyon},\ and\ \citenamefont
  {Yoshimura}}]{CastroAlvaredo}%
  \BibitemOpen
  \bibfield  {author} {\bibinfo {author} {\bibfnamefont {O.~A.}\ \bibnamefont
  {Castro-Alvaredo}}, \bibinfo {author} {\bibfnamefont {B.}~\bibnamefont
  {Doyon}},\ and\ \bibinfo {author} {\bibfnamefont {T.}~\bibnamefont
  {Yoshimura}},\ }\bibfield  {title} {\bibinfo {title} {Emergent hydrodynamics
  in integrable quantum systems out of equilibrium},\ }\href
  {https://doi.org/10.1103/PhysRevX.6.041065} {\bibfield  {journal} {\bibinfo
  {journal} {Phys. Rev. X}\ }\textbf {\bibinfo {volume} {6}},\ \bibinfo {pages}
  {041065} (\bibinfo {year} {2016})}\BibitemShut {NoStop}%
\bibitem [{\citenamefont {Bertini}\ \emph {et~al.}(2016)\citenamefont
  {Bertini}, \citenamefont {Collura}, \citenamefont {De~Nardis},\ and\
  \citenamefont {Fagotti}}]{Fagotti}%
  \BibitemOpen
  \bibfield  {author} {\bibinfo {author} {\bibfnamefont {B.}~\bibnamefont
  {Bertini}}, \bibinfo {author} {\bibfnamefont {M.}~\bibnamefont {Collura}},
  \bibinfo {author} {\bibfnamefont {J.}~\bibnamefont {De~Nardis}},\ and\
  \bibinfo {author} {\bibfnamefont {M.}~\bibnamefont {Fagotti}},\ }\bibfield
  {title} {\bibinfo {title} {{Transport in Out-of-Equilibrium $XXZ$ Chains:
  Exact Profiles of Charges and Currents}},\ }\href
  {https://doi.org/10.1103/PhysRevLett.117.207201} {\bibfield  {journal}
  {\bibinfo  {journal} {Phys. Rev. Lett.}\ }\textbf {\bibinfo {volume} {117}},\
  \bibinfo {pages} {207201} (\bibinfo {year} {2016})}\BibitemShut {NoStop}%
\bibitem [{\citenamefont {Bulchandani}\ \emph {et~al.}(2018)\citenamefont
  {Bulchandani}, \citenamefont {Vasseur}, \citenamefont {Karrasch},\ and\
  \citenamefont {Moore}}]{BVKM1}%
  \BibitemOpen
  \bibfield  {author} {\bibinfo {author} {\bibfnamefont {V.~B.}\ \bibnamefont
  {Bulchandani}}, \bibinfo {author} {\bibfnamefont {R.}~\bibnamefont
  {Vasseur}}, \bibinfo {author} {\bibfnamefont {C.}~\bibnamefont {Karrasch}},\
  and\ \bibinfo {author} {\bibfnamefont {J.~E.}\ \bibnamefont {Moore}},\
  }\bibfield  {title} {\bibinfo {title} {{Bethe-Boltzmann hydrodynamics and
  spin transport in the XXZ chain}},\ }\bibfield  {journal} {\bibinfo
  {journal} {Physical Review B}\ }\textbf {\bibinfo {volume} {97}},\ \href
  {https://doi.org/10.1103/PhysRevB.97.045407} {10.1103/PhysRevB.97.045407}
  (\bibinfo {year} {2018}),\ \Eprint {https://arxiv.org/abs/1702.06146}
  {arXiv:1702.06146} \BibitemShut {NoStop}%
\bibitem [{\citenamefont {Doyon}\ \emph {et~al.}(2018)\citenamefont {Doyon},
  \citenamefont {Yoshimura},\ and\ \citenamefont {Caux}}]{DoyonSoliton}%
  \BibitemOpen
  \bibfield  {author} {\bibinfo {author} {\bibfnamefont {B.}~\bibnamefont
  {Doyon}}, \bibinfo {author} {\bibfnamefont {T.}~\bibnamefont {Yoshimura}},\
  and\ \bibinfo {author} {\bibfnamefont {J.-S.}\ \bibnamefont {Caux}},\
  }\bibfield  {title} {\bibinfo {title} {Soliton gases and generalized
  hydrodynamics},\ }\href {https://doi.org/10.1103/PhysRevLett.120.045301}
  {\bibfield  {journal} {\bibinfo  {journal} {Phys. Rev. Lett.}\ }\textbf
  {\bibinfo {volume} {120}},\ \bibinfo {pages} {045301} (\bibinfo {year}
  {2018})}\BibitemShut {NoStop}%
\bibitem [{\citenamefont {Bastianello}\ \emph {et~al.}(2020)\citenamefont
  {Bastianello}, \citenamefont {De~Nardis},\ and\ \citenamefont
  {De~Luca}}]{DephNoise}%
  \BibitemOpen
  \bibfield  {author} {\bibinfo {author} {\bibfnamefont {A.}~\bibnamefont
  {Bastianello}}, \bibinfo {author} {\bibfnamefont {J.}~\bibnamefont
  {De~Nardis}},\ and\ \bibinfo {author} {\bibfnamefont {A.}~\bibnamefont
  {De~Luca}},\ }\bibfield  {title} {\bibinfo {title} {Generalized hydrodynamics
  with dephasing noise},\ }\href {https://doi.org/10.1103/PhysRevB.102.161110}
  {\bibfield  {journal} {\bibinfo  {journal} {Phys. Rev. B}\ }\textbf {\bibinfo
  {volume} {102}},\ \bibinfo {pages} {161110} (\bibinfo {year}
  {2020})}\BibitemShut {NoStop}%
\bibitem [{\citenamefont {Lifshitz}\ and\ \citenamefont
  {Pitaevskii}(2013)}]{lifshitz2013statistical}%
  \BibitemOpen
  \bibfield  {author} {\bibinfo {author} {\bibfnamefont {E.~M.}\ \bibnamefont
  {Lifshitz}}\ and\ \bibinfo {author} {\bibfnamefont {L.~P.}\ \bibnamefont
  {Pitaevskii}},\ }\href@noop {} {\emph {\bibinfo {title} {Statistical physics:
  theory of the condensed state}}},\ Vol.~\bibinfo {volume} {9}\ (\bibinfo
  {publisher} {Elsevier},\ \bibinfo {year} {2013})\BibitemShut {NoStop}%
\bibitem [{\citenamefont {Kiely}\ and\ \citenamefont
  {Mueller}(2021)}]{Mueller}%
  \BibitemOpen
  \bibfield  {author} {\bibinfo {author} {\bibfnamefont {T.~G.}\ \bibnamefont
  {Kiely}}\ and\ \bibinfo {author} {\bibfnamefont {E.~J.}\ \bibnamefont
  {Mueller}},\ }\bibfield  {title} {\bibinfo {title} {{Transport in the
  two-dimensional Fermi-Hubbard model: Lessons from weak coupling}},\ }\href
  {https://doi.org/10.1103/PhysRevB.104.165143} {\bibfield  {journal} {\bibinfo
   {journal} {Phys. Rev. B}\ }\textbf {\bibinfo {volume} {104}},\ \bibinfo
  {pages} {165143} (\bibinfo {year} {2021})}\BibitemShut {NoStop}%
\bibitem [{\citenamefont {Vu\ifmmode \check{c}\else \v{c}\fi{}i\ifmmode
  \check{c}\else \v{c}\fi{}evi\ifmmode~\acute{c}\else \'{c}\fi{}}\ \emph
  {et~al.}(2023)\citenamefont {Vu\ifmmode \check{c}\else \v{c}\fi{}i\ifmmode
  \check{c}\else \v{c}\fi{}evi\ifmmode~\acute{c}\else \'{c}\fi{}},
  \citenamefont {Predin},\ and\ \citenamefont {Ferrero}}]{Vucicevic}%
  \BibitemOpen
  \bibfield  {author} {\bibinfo {author} {\bibfnamefont {J.}~\bibnamefont
  {Vu\ifmmode \check{c}\else \v{c}\fi{}i\ifmmode \check{c}\else
  \v{c}\fi{}evi\ifmmode~\acute{c}\else \'{c}\fi{}}}, \bibinfo {author}
  {\bibfnamefont {S.}~\bibnamefont {Predin}},\ and\ \bibinfo {author}
  {\bibfnamefont {M.}~\bibnamefont {Ferrero}},\ }\bibfield  {title} {\bibinfo
  {title} {{Charge fluctuations, hydrodynamics, and transport in the
  square-lattice Hubbard model}},\ }\href
  {https://doi.org/10.1103/PhysRevB.107.155140} {\bibfield  {journal} {\bibinfo
   {journal} {Phys. Rev. B}\ }\textbf {\bibinfo {volume} {107}},\ \bibinfo
  {pages} {155140} (\bibinfo {year} {2023})}\BibitemShut {NoStop}%
\bibitem [{\citenamefont {Grozdanov}\ \emph {et~al.}(2019)\citenamefont
  {Grozdanov}, \citenamefont {Lucas},\ and\ \citenamefont
  {Poovuttikul}}]{Quasihydro}%
  \BibitemOpen
  \bibfield  {author} {\bibinfo {author} {\bibfnamefont {S.}~\bibnamefont
  {Grozdanov}}, \bibinfo {author} {\bibfnamefont {A.}~\bibnamefont {Lucas}},\
  and\ \bibinfo {author} {\bibfnamefont {N.}~\bibnamefont {Poovuttikul}},\
  }\bibfield  {title} {\bibinfo {title} {Holography and hydrodynamics with
  weakly broken symmetries},\ }\href
  {https://doi.org/10.1103/PhysRevD.99.086012} {\bibfield  {journal} {\bibinfo
  {journal} {Phys. Rev. D}\ }\textbf {\bibinfo {volume} {99}},\ \bibinfo
  {pages} {086012} (\bibinfo {year} {2019})}\BibitemShut {NoStop}%
\bibitem [{Sup()}]{SuppMat}%
  \BibitemOpen
  \href@noop {} {\bibinfo {title} {{See Appendices for further
  details.}}}\BibitemShut {Stop}%
\bibitem [{\citenamefont {Mahan}(2013)}]{mahan2013many}%
  \BibitemOpen
  \bibfield  {author} {\bibinfo {author} {\bibfnamefont {G.~D.}\ \bibnamefont
  {Mahan}},\ }\href@noop {} {\emph {\bibinfo {title} {Many-particle physics}}}\
  (\bibinfo  {publisher} {Springer Science \& Business Media},\ \bibinfo {year}
  {2013})\BibitemShut {NoStop}%
\bibitem [{\citenamefont {Kim}\ \emph {et~al.}(2015)\citenamefont {Kim},
  \citenamefont {Bañuls}, \citenamefont {Cirac}, \citenamefont {Hastings},\
  and\ \citenamefont {Huse}}]{Kim_2015}%
  \BibitemOpen
  \bibfield  {author} {\bibinfo {author} {\bibfnamefont {H.}~\bibnamefont
  {Kim}}, \bibinfo {author} {\bibfnamefont {M.~C.}\ \bibnamefont {Bañuls}},
  \bibinfo {author} {\bibfnamefont {J.~I.}\ \bibnamefont {Cirac}}, \bibinfo
  {author} {\bibfnamefont {M.~B.}\ \bibnamefont {Hastings}},\ and\ \bibinfo
  {author} {\bibfnamefont {D.~A.}\ \bibnamefont {Huse}},\ }\bibfield  {title}
  {\bibinfo {title} {Slowest local operators in quantum spin chains},\
  }\bibfield  {journal} {\bibinfo  {journal} {Physical Review E}\ }\textbf
  {\bibinfo {volume} {92}},\ \href {https://doi.org/10.1103/physreve.92.012128}
  {10.1103/physreve.92.012128} (\bibinfo {year} {2015})\BibitemShut {NoStop}%
\bibitem [{Note1()}]{Note1}%
  \BibitemOpen
  \bibinfo {note} {This model in fact exhibits superdiffusive charge
  transport\cite
  {Znidaric,SuperdiffLattice,scheie2021detection,Blochexp,Bulchandani_2021}.
  Tuning the effective XXZ anisotropy $\Delta =U_1$ of the integrable reference
  point in our constrained optimization problem is equivalent to tuning the
  value of the interaction constraint $\sigma _V^2$. We have checked that even
  when $\sigma _V^2/L<1/4$, which corresponds to an anisotropy $|\Delta | <1$
  and hence ballistic charge transport for the integrable reference point, the
  optimal models with $R>1$ continue to exhibit a clear short-time enhancement
  of band sound compared to the integrable reference point with $R=1$. One
  interesting corollary of our analysis with $\Delta =1$ is that spin
  superdiffusion in the spin-$1/2$ Heisenberg chain occurs \protect \emph
  {despite} the presence of substantial Umklapp scattering.}\BibitemShut
  {Stop}%
\bibitem [{\citenamefont {Oganesyan}\ and\ \citenamefont {Huse}(2007)}]{rstat}%
  \BibitemOpen
  \bibfield  {author} {\bibinfo {author} {\bibfnamefont {V.}~\bibnamefont
  {Oganesyan}}\ and\ \bibinfo {author} {\bibfnamefont {D.~A.}\ \bibnamefont
  {Huse}},\ }\bibfield  {title} {\bibinfo {title} {Localization of interacting
  fermions at high temperature},\ }\href
  {https://doi.org/10.1103/PhysRevB.75.155111} {\bibfield  {journal} {\bibinfo
  {journal} {Phys. Rev. B}\ }\textbf {\bibinfo {volume} {75}},\ \bibinfo
  {pages} {155111} (\bibinfo {year} {2007})}\BibitemShut {NoStop}%
\bibitem [{\citenamefont {Ledwith}\ \emph {et~al.}(2019)\citenamefont
  {Ledwith}, \citenamefont {Guo},\ and\ \citenamefont
  {Levitov}}]{Ledwith_2019}%
  \BibitemOpen
  \bibfield  {author} {\bibinfo {author} {\bibfnamefont {P.~J.}\ \bibnamefont
  {Ledwith}}, \bibinfo {author} {\bibfnamefont {H.}~\bibnamefont {Guo}},\ and\
  \bibinfo {author} {\bibfnamefont {L.}~\bibnamefont {Levitov}},\ }\bibfield
  {title} {\bibinfo {title} {{The hierarchy of excitation lifetimes in
  two-dimensional Fermi gases}},\ }\href
  {https://doi.org/10.1016/j.aop.2019.167913} {\bibfield  {journal} {\bibinfo
  {journal} {Annals of Physics}\ }\textbf {\bibinfo {volume} {411}},\ \bibinfo
  {pages} {167913} (\bibinfo {year} {2019})}\BibitemShut {NoStop}%
\bibitem [{\citenamefont {Monroe}\ \emph {et~al.}(2021)\citenamefont {Monroe},
  \citenamefont {Campbell}, \citenamefont {Duan}, \citenamefont {Gong},
  \citenamefont {Gorshkov}, \citenamefont {Hess}, \citenamefont {Islam},
  \citenamefont {Kim}, \citenamefont {Linke}, \citenamefont {Pagano},
  \citenamefont {Richerme}, \citenamefont {Senko},\ and\ \citenamefont
  {Yao}}]{trapped_ion_review}%
  \BibitemOpen
  \bibfield  {author} {\bibinfo {author} {\bibfnamefont {C.}~\bibnamefont
  {Monroe}}, \bibinfo {author} {\bibfnamefont {W.~C.}\ \bibnamefont
  {Campbell}}, \bibinfo {author} {\bibfnamefont {L.-M.}\ \bibnamefont {Duan}},
  \bibinfo {author} {\bibfnamefont {Z.-X.}\ \bibnamefont {Gong}}, \bibinfo
  {author} {\bibfnamefont {A.~V.}\ \bibnamefont {Gorshkov}}, \bibinfo {author}
  {\bibfnamefont {P.~W.}\ \bibnamefont {Hess}}, \bibinfo {author}
  {\bibfnamefont {R.}~\bibnamefont {Islam}}, \bibinfo {author} {\bibfnamefont
  {K.}~\bibnamefont {Kim}}, \bibinfo {author} {\bibfnamefont {N.~M.}\
  \bibnamefont {Linke}}, \bibinfo {author} {\bibfnamefont {G.}~\bibnamefont
  {Pagano}}, \bibinfo {author} {\bibfnamefont {P.}~\bibnamefont {Richerme}},
  \bibinfo {author} {\bibfnamefont {C.}~\bibnamefont {Senko}},\ and\ \bibinfo
  {author} {\bibfnamefont {N.~Y.}\ \bibnamefont {Yao}},\ }\bibfield  {title}
  {\bibinfo {title} {Programmable quantum simulations of spin systems with
  trapped ions},\ }\href {https://doi.org/10.1103/RevModPhys.93.025001}
  {\bibfield  {journal} {\bibinfo  {journal} {Rev. Mod. Phys.}\ }\textbf
  {\bibinfo {volume} {93}},\ \bibinfo {pages} {025001} (\bibinfo {year}
  {2021})}\BibitemShut {NoStop}%
\bibitem [{\citenamefont {Korenblit}\ \emph {et~al.}(2012)\citenamefont
  {Korenblit}, \citenamefont {Kafri}, \citenamefont {Campbell}, \citenamefont
  {Islam}, \citenamefont {Edwards}, \citenamefont {Gong}, \citenamefont {Lin},
  \citenamefont {Duan}, \citenamefont {Kim}, \citenamefont {Kim},\ and\
  \citenamefont {Monroe}}]{Tunable}%
  \BibitemOpen
  \bibfield  {author} {\bibinfo {author} {\bibfnamefont {S.}~\bibnamefont
  {Korenblit}}, \bibinfo {author} {\bibfnamefont {D.}~\bibnamefont {Kafri}},
  \bibinfo {author} {\bibfnamefont {W.~C.}\ \bibnamefont {Campbell}}, \bibinfo
  {author} {\bibfnamefont {R.}~\bibnamefont {Islam}}, \bibinfo {author}
  {\bibfnamefont {E.~E.}\ \bibnamefont {Edwards}}, \bibinfo {author}
  {\bibfnamefont {Z.-X.}\ \bibnamefont {Gong}}, \bibinfo {author}
  {\bibfnamefont {G.-D.}\ \bibnamefont {Lin}}, \bibinfo {author} {\bibfnamefont
  {L.-M.}\ \bibnamefont {Duan}}, \bibinfo {author} {\bibfnamefont
  {J.}~\bibnamefont {Kim}}, \bibinfo {author} {\bibfnamefont {K.}~\bibnamefont
  {Kim}},\ and\ \bibinfo {author} {\bibfnamefont {C.}~\bibnamefont {Monroe}},\
  }\bibfield  {title} {\bibinfo {title} {Quantum simulation of spin models on
  an arbitrary lattice with trapped ions},\ }\href
  {https://doi.org/10.1088/1367-2630/14/9/095024} {\bibfield  {journal}
  {\bibinfo  {journal} {New Journal of Physics}\ }\textbf {\bibinfo {volume}
  {14}},\ \bibinfo {pages} {095024} (\bibinfo {year} {2012})}\BibitemShut
  {NoStop}%
\bibitem [{\citenamefont {Moses}\ \emph {et~al.}(2023)\citenamefont {Moses},
  \citenamefont {Baldwin}, \citenamefont {Allman}, \citenamefont {Ancona},
  \citenamefont {Ascarrunz}, \citenamefont {Barnes}, \citenamefont
  {Bartolotta}, \citenamefont {Bjork}, \citenamefont {Blanchard}, \citenamefont
  {Bohn}, \citenamefont {Bohnet}, \citenamefont {Brown}, \citenamefont
  {Burdick}, \citenamefont {Burton}, \citenamefont {Campbell}, \citenamefont
  {Campora}, \citenamefont {Carron}, \citenamefont {Chambers}, \citenamefont
  {Chan}, \citenamefont {Chen}, \citenamefont {Chernoguzov}, \citenamefont
  {Chertkov}, \citenamefont {Colina}, \citenamefont {Curtis}, \citenamefont
  {Daniel}, \citenamefont {DeCross}, \citenamefont {Deen}, \citenamefont
  {Delaney}, \citenamefont {Dreiling}, \citenamefont {Ertsgaard}, \citenamefont
  {Esposito}, \citenamefont {Estey}, \citenamefont {Fabrikant}, \citenamefont
  {Figgatt}, \citenamefont {Foltz}, \citenamefont {Foss-Feig}, \citenamefont
  {Francois}, \citenamefont {Gaebler}, \citenamefont {Gatterman}, \citenamefont
  {Gilbreth}, \citenamefont {Giles}, \citenamefont {Glynn}, \citenamefont
  {Hall}, \citenamefont {Hankin}, \citenamefont {Hansen}, \citenamefont
  {Hayes}, \citenamefont {Higashi}, \citenamefont {Hoffman}, \citenamefont
  {Horning}, \citenamefont {Hout}, \citenamefont {Jacobs}, \citenamefont
  {Johansen}, \citenamefont {Jones}, \citenamefont {Karcz}, \citenamefont
  {Klein}, \citenamefont {Lauria}, \citenamefont {Lee}, \citenamefont {Liefer},
  \citenamefont {Lu}, \citenamefont {Lucchetti}, \citenamefont {Lytle},
  \citenamefont {Malm}, \citenamefont {Matheny}, \citenamefont {Mathewson},
  \citenamefont {Mayer}, \citenamefont {Miller}, \citenamefont {Mills},
  \citenamefont {Neyenhuis}, \citenamefont {Nugent}, \citenamefont {Olson},
  \citenamefont {Parks}, \citenamefont {Price}, \citenamefont {Price},
  \citenamefont {Pugh}, \citenamefont {Ransford}, \citenamefont {Reed},
  \citenamefont {Roman}, \citenamefont {Rowe}, \citenamefont {Ryan-Anderson},
  \citenamefont {Sanders}, \citenamefont {Sedlacek}, \citenamefont {Shevchuk},
  \citenamefont {Siegfried}, \citenamefont {Skripka}, \citenamefont {Spaun},
  \citenamefont {Sprenkle}, \citenamefont {Stutz}, \citenamefont {Swallows},
  \citenamefont {Tobey}, \citenamefont {Tran}, \citenamefont {Tran},
  \citenamefont {Vogt}, \citenamefont {Volin}, \citenamefont {Walker},
  \citenamefont {Zolot},\ and\ \citenamefont {Pino}}]{PhysRevX.13.041052}%
  \BibitemOpen
  \bibfield  {author} {\bibinfo {author} {\bibfnamefont {S.~A.}\ \bibnamefont
  {Moses}}, \bibinfo {author} {\bibfnamefont {C.~H.}\ \bibnamefont {Baldwin}},
  \bibinfo {author} {\bibfnamefont {M.~S.}\ \bibnamefont {Allman}}, \bibinfo
  {author} {\bibfnamefont {R.}~\bibnamefont {Ancona}}, \bibinfo {author}
  {\bibfnamefont {L.}~\bibnamefont {Ascarrunz}}, \bibinfo {author}
  {\bibfnamefont {C.}~\bibnamefont {Barnes}}, \bibinfo {author} {\bibfnamefont
  {J.}~\bibnamefont {Bartolotta}}, \bibinfo {author} {\bibfnamefont
  {B.}~\bibnamefont {Bjork}}, \bibinfo {author} {\bibfnamefont
  {P.}~\bibnamefont {Blanchard}}, \bibinfo {author} {\bibfnamefont
  {M.}~\bibnamefont {Bohn}}, \bibinfo {author} {\bibfnamefont {J.~G.}\
  \bibnamefont {Bohnet}}, \bibinfo {author} {\bibfnamefont {N.~C.}\
  \bibnamefont {Brown}}, \bibinfo {author} {\bibfnamefont {N.~Q.}\ \bibnamefont
  {Burdick}}, \bibinfo {author} {\bibfnamefont {W.~C.}\ \bibnamefont {Burton}},
  \bibinfo {author} {\bibfnamefont {S.~L.}\ \bibnamefont {Campbell}}, \bibinfo
  {author} {\bibfnamefont {J.~P.}\ \bibnamefont {Campora}}, \bibinfo {author}
  {\bibfnamefont {C.}~\bibnamefont {Carron}}, \bibinfo {author} {\bibfnamefont
  {J.}~\bibnamefont {Chambers}}, \bibinfo {author} {\bibfnamefont {J.~W.}\
  \bibnamefont {Chan}}, \bibinfo {author} {\bibfnamefont {Y.~H.}\ \bibnamefont
  {Chen}}, \bibinfo {author} {\bibfnamefont {A.}~\bibnamefont {Chernoguzov}},
  \bibinfo {author} {\bibfnamefont {E.}~\bibnamefont {Chertkov}}, \bibinfo
  {author} {\bibfnamefont {J.}~\bibnamefont {Colina}}, \bibinfo {author}
  {\bibfnamefont {J.~P.}\ \bibnamefont {Curtis}}, \bibinfo {author}
  {\bibfnamefont {R.}~\bibnamefont {Daniel}}, \bibinfo {author} {\bibfnamefont
  {M.}~\bibnamefont {DeCross}}, \bibinfo {author} {\bibfnamefont
  {D.}~\bibnamefont {Deen}}, \bibinfo {author} {\bibfnamefont {C.}~\bibnamefont
  {Delaney}}, \bibinfo {author} {\bibfnamefont {J.~M.}\ \bibnamefont
  {Dreiling}}, \bibinfo {author} {\bibfnamefont {C.~T.}\ \bibnamefont
  {Ertsgaard}}, \bibinfo {author} {\bibfnamefont {J.}~\bibnamefont {Esposito}},
  \bibinfo {author} {\bibfnamefont {B.}~\bibnamefont {Estey}}, \bibinfo
  {author} {\bibfnamefont {M.}~\bibnamefont {Fabrikant}}, \bibinfo {author}
  {\bibfnamefont {C.}~\bibnamefont {Figgatt}}, \bibinfo {author} {\bibfnamefont
  {C.}~\bibnamefont {Foltz}}, \bibinfo {author} {\bibfnamefont
  {M.}~\bibnamefont {Foss-Feig}}, \bibinfo {author} {\bibfnamefont
  {D.}~\bibnamefont {Francois}}, \bibinfo {author} {\bibfnamefont {J.~P.}\
  \bibnamefont {Gaebler}}, \bibinfo {author} {\bibfnamefont {T.~M.}\
  \bibnamefont {Gatterman}}, \bibinfo {author} {\bibfnamefont {C.~N.}\
  \bibnamefont {Gilbreth}}, \bibinfo {author} {\bibfnamefont {J.}~\bibnamefont
  {Giles}}, \bibinfo {author} {\bibfnamefont {E.}~\bibnamefont {Glynn}},
  \bibinfo {author} {\bibfnamefont {A.}~\bibnamefont {Hall}}, \bibinfo {author}
  {\bibfnamefont {A.~M.}\ \bibnamefont {Hankin}}, \bibinfo {author}
  {\bibfnamefont {A.}~\bibnamefont {Hansen}}, \bibinfo {author} {\bibfnamefont
  {D.}~\bibnamefont {Hayes}}, \bibinfo {author} {\bibfnamefont
  {B.}~\bibnamefont {Higashi}}, \bibinfo {author} {\bibfnamefont {I.~M.}\
  \bibnamefont {Hoffman}}, \bibinfo {author} {\bibfnamefont {B.}~\bibnamefont
  {Horning}}, \bibinfo {author} {\bibfnamefont {J.~J.}\ \bibnamefont {Hout}},
  \bibinfo {author} {\bibfnamefont {R.}~\bibnamefont {Jacobs}}, \bibinfo
  {author} {\bibfnamefont {J.}~\bibnamefont {Johansen}}, \bibinfo {author}
  {\bibfnamefont {L.}~\bibnamefont {Jones}}, \bibinfo {author} {\bibfnamefont
  {J.}~\bibnamefont {Karcz}}, \bibinfo {author} {\bibfnamefont
  {T.}~\bibnamefont {Klein}}, \bibinfo {author} {\bibfnamefont
  {P.}~\bibnamefont {Lauria}}, \bibinfo {author} {\bibfnamefont
  {P.}~\bibnamefont {Lee}}, \bibinfo {author} {\bibfnamefont {D.}~\bibnamefont
  {Liefer}}, \bibinfo {author} {\bibfnamefont {S.~T.}\ \bibnamefont {Lu}},
  \bibinfo {author} {\bibfnamefont {D.}~\bibnamefont {Lucchetti}}, \bibinfo
  {author} {\bibfnamefont {C.}~\bibnamefont {Lytle}}, \bibinfo {author}
  {\bibfnamefont {A.}~\bibnamefont {Malm}}, \bibinfo {author} {\bibfnamefont
  {M.}~\bibnamefont {Matheny}}, \bibinfo {author} {\bibfnamefont
  {B.}~\bibnamefont {Mathewson}}, \bibinfo {author} {\bibfnamefont
  {K.}~\bibnamefont {Mayer}}, \bibinfo {author} {\bibfnamefont {D.~B.}\
  \bibnamefont {Miller}}, \bibinfo {author} {\bibfnamefont {M.}~\bibnamefont
  {Mills}}, \bibinfo {author} {\bibfnamefont {B.}~\bibnamefont {Neyenhuis}},
  \bibinfo {author} {\bibfnamefont {L.}~\bibnamefont {Nugent}}, \bibinfo
  {author} {\bibfnamefont {S.}~\bibnamefont {Olson}}, \bibinfo {author}
  {\bibfnamefont {J.}~\bibnamefont {Parks}}, \bibinfo {author} {\bibfnamefont
  {G.~N.}\ \bibnamefont {Price}}, \bibinfo {author} {\bibfnamefont
  {Z.}~\bibnamefont {Price}}, \bibinfo {author} {\bibfnamefont
  {M.}~\bibnamefont {Pugh}}, \bibinfo {author} {\bibfnamefont {A.}~\bibnamefont
  {Ransford}}, \bibinfo {author} {\bibfnamefont {A.~P.}\ \bibnamefont {Reed}},
  \bibinfo {author} {\bibfnamefont {C.}~\bibnamefont {Roman}}, \bibinfo
  {author} {\bibfnamefont {M.}~\bibnamefont {Rowe}}, \bibinfo {author}
  {\bibfnamefont {C.}~\bibnamefont {Ryan-Anderson}}, \bibinfo {author}
  {\bibfnamefont {S.}~\bibnamefont {Sanders}}, \bibinfo {author} {\bibfnamefont
  {J.}~\bibnamefont {Sedlacek}}, \bibinfo {author} {\bibfnamefont
  {P.}~\bibnamefont {Shevchuk}}, \bibinfo {author} {\bibfnamefont
  {P.}~\bibnamefont {Siegfried}}, \bibinfo {author} {\bibfnamefont
  {T.}~\bibnamefont {Skripka}}, \bibinfo {author} {\bibfnamefont
  {B.}~\bibnamefont {Spaun}}, \bibinfo {author} {\bibfnamefont {R.~T.}\
  \bibnamefont {Sprenkle}}, \bibinfo {author} {\bibfnamefont {R.~P.}\
  \bibnamefont {Stutz}}, \bibinfo {author} {\bibfnamefont {M.}~\bibnamefont
  {Swallows}}, \bibinfo {author} {\bibfnamefont {R.~I.}\ \bibnamefont {Tobey}},
  \bibinfo {author} {\bibfnamefont {A.}~\bibnamefont {Tran}}, \bibinfo {author}
  {\bibfnamefont {T.}~\bibnamefont {Tran}}, \bibinfo {author} {\bibfnamefont
  {E.}~\bibnamefont {Vogt}}, \bibinfo {author} {\bibfnamefont {C.}~\bibnamefont
  {Volin}}, \bibinfo {author} {\bibfnamefont {J.}~\bibnamefont {Walker}},
  \bibinfo {author} {\bibfnamefont {A.~M.}\ \bibnamefont {Zolot}},\ and\
  \bibinfo {author} {\bibfnamefont {J.~M.}\ \bibnamefont {Pino}},\ }\bibfield
  {title} {\bibinfo {title} {A race-track trapped-ion quantum processor},\
  }\href {https://doi.org/10.1103/PhysRevX.13.041052} {\bibfield  {journal}
  {\bibinfo  {journal} {Phys. Rev. X}\ }\textbf {\bibinfo {volume} {13}},\
  \bibinfo {pages} {041052} (\bibinfo {year} {2023})}\BibitemShut {NoStop}%
\bibitem [{\citenamefont {Andersen}\ \emph {et~al.}(2024)\citenamefont
  {Andersen}, \citenamefont {Hoke}, \citenamefont {Drozdov}, \citenamefont
  {Klimov} \emph {et~al.}}]{rosenberg2024dynamics}%
  \BibitemOpen
  \bibfield  {author} {\bibinfo {author} {\bibfnamefont {T.}~\bibnamefont
  {Andersen}}, \bibinfo {author} {\bibfnamefont {J.}~\bibnamefont {Hoke}},
  \bibinfo {author} {\bibfnamefont {I.}~\bibnamefont {Drozdov}}, \bibinfo
  {author} {\bibfnamefont {P.}~\bibnamefont {Klimov}}, \emph {et~al.},\
  }\bibfield  {title} {\bibinfo {title} {{Dynamics of magnetization at infinite
  temperature in a Heisenberg spin chain}},\ }\href@noop {} {\bibfield
  {journal} {\bibinfo  {journal} {Science}\ }\textbf {\bibinfo {volume}
  {384}},\ \bibinfo {pages} {48} (\bibinfo {year} {2024})}\BibitemShut
  {NoStop}%
\bibitem [{\citenamefont {Mierzejewski}\ \emph {et~al.}(2023)\citenamefont
  {Mierzejewski}, \citenamefont {Wronowicz}, \citenamefont {Paw\l{}owski},\
  and\ \citenamefont {Herbrych}}]{Quasiballistic}%
  \BibitemOpen
  \bibfield  {author} {\bibinfo {author} {\bibfnamefont {M.}~\bibnamefont
  {Mierzejewski}}, \bibinfo {author} {\bibfnamefont {J.}~\bibnamefont
  {Wronowicz}}, \bibinfo {author} {\bibfnamefont {J.}~\bibnamefont
  {Paw\l{}owski}},\ and\ \bibinfo {author} {\bibfnamefont {J.}~\bibnamefont
  {Herbrych}},\ }\bibfield  {title} {\bibinfo {title} {{Quasiballistic
  transport in the long-range anisotropic Heisenberg model}},\ }\href
  {https://doi.org/10.1103/PhysRevB.107.045134} {\bibfield  {journal} {\bibinfo
   {journal} {Phys. Rev. B}\ }\textbf {\bibinfo {volume} {107}},\ \bibinfo
  {pages} {045134} (\bibinfo {year} {2023})}\BibitemShut {NoStop}%
\bibitem [{\citenamefont {Yang}\ \emph {et~al.}(2024)\citenamefont {Yang},
  \citenamefont {Ma},\ and\ \citenamefont {Ying}}]{PhysRevB.110.014308}%
  \BibitemOpen
  \bibfield  {author} {\bibinfo {author} {\bibfnamefont {A.}~\bibnamefont
  {Yang}}, \bibinfo {author} {\bibfnamefont {J.-L.}\ \bibnamefont {Ma}},\ and\
  \bibinfo {author} {\bibfnamefont {L.}~\bibnamefont {Ying}},\ }\bibfield
  {title} {\bibinfo {title} {{Emergent anomalous hydrodynamics at infinite
  temperature in a long-range XXZ model}},\ }\href
  {https://doi.org/10.1103/PhysRevB.110.014308} {\bibfield  {journal} {\bibinfo
   {journal} {Phys. Rev. B}\ }\textbf {\bibinfo {volume} {110}},\ \bibinfo
  {pages} {014308} (\bibinfo {year} {2024})}\BibitemShut {NoStop}%
\bibitem [{\citenamefont {Song}\ \emph {et~al.}(2025)\citenamefont {Song},
  \citenamefont {Ha}, \citenamefont {Ho},\ and\ \citenamefont
  {Bulchandani}}]{khpy-d9s7}%
  \BibitemOpen
  \bibfield  {author} {\bibinfo {author} {\bibfnamefont {J.~Z.}\ \bibnamefont
  {Song}}, \bibinfo {author} {\bibfnamefont {H.}~\bibnamefont {Ha}}, \bibinfo
  {author} {\bibfnamefont {W.~W.}\ \bibnamefont {Ho}},\ and\ \bibinfo {author}
  {\bibfnamefont {V.~B.}\ \bibnamefont {Bulchandani}},\ }\bibfield  {title}
  {\bibinfo {title} {Theory of quasiballistic spin transport},\ }\href
  {https://doi.org/10.1103/khpy-d9s7} {\bibfield  {journal} {\bibinfo
  {journal} {Phys. Rev. B}\ }\textbf {\bibinfo {volume} {112}},\ \bibinfo
  {pages} {L241408} (\bibinfo {year} {2025})}\BibitemShut {NoStop}%
\bibitem [{\citenamefont {Ljubotina}\ \emph {et~al.}(2023)\citenamefont
  {Ljubotina}, \citenamefont {Desaules}, \citenamefont {Serbyn},\ and\
  \citenamefont {Papić}}]{Ljubotina_2023}%
  \BibitemOpen
  \bibfield  {author} {\bibinfo {author} {\bibfnamefont {M.}~\bibnamefont
  {Ljubotina}}, \bibinfo {author} {\bibfnamefont {J.-Y.}\ \bibnamefont
  {Desaules}}, \bibinfo {author} {\bibfnamefont {M.}~\bibnamefont {Serbyn}},\
  and\ \bibinfo {author} {\bibfnamefont {Z.}~\bibnamefont {Papić}},\
  }\bibfield  {title} {\bibinfo {title} {Superdiffusive energy transport in
  kinetically constrained models},\ }\bibfield  {journal} {\bibinfo  {journal}
  {Physical Review X}\ }\textbf {\bibinfo {volume} {13}},\ \href
  {https://doi.org/10.1103/physrevx.13.011033} {10.1103/physrevx.13.011033}
  (\bibinfo {year} {2023})\BibitemShut {NoStop}%
\bibitem [{\citenamefont {Yi-Thomas}\ \emph {et~al.}(2024)\citenamefont
  {Yi-Thomas}, \citenamefont {Ware}, \citenamefont {Sau},\ and\ \citenamefont
  {White}}]{PhysRevB.110.134308}%
  \BibitemOpen
  \bibfield  {author} {\bibinfo {author} {\bibfnamefont {S.}~\bibnamefont
  {Yi-Thomas}}, \bibinfo {author} {\bibfnamefont {B.}~\bibnamefont {Ware}},
  \bibinfo {author} {\bibfnamefont {J.~D.}\ \bibnamefont {Sau}},\ and\ \bibinfo
  {author} {\bibfnamefont {C.~D.}\ \bibnamefont {White}},\ }\bibfield  {title}
  {\bibinfo {title} {Comparing numerical methods for hydrodynamics in a
  one-dimensional lattice spin model},\ }\href
  {https://doi.org/10.1103/PhysRevB.110.134308} {\bibfield  {journal} {\bibinfo
   {journal} {Phys. Rev. B}\ }\textbf {\bibinfo {volume} {110}},\ \bibinfo
  {pages} {134308} (\bibinfo {year} {2024})}\BibitemShut {NoStop}%
\bibitem [{\citenamefont {Mark}\ \emph {et~al.}(2025)\citenamefont {Mark},
  \citenamefont {Surace}, \citenamefont {Schuster}, \citenamefont {Shaw},
  \citenamefont {Gong}, \citenamefont {Choi},\ and\ \citenamefont
  {Endres}}]{mark2025observationballisticplasmamemory}%
  \BibitemOpen
  \bibfield  {author} {\bibinfo {author} {\bibfnamefont {D.~K.}\ \bibnamefont
  {Mark}}, \bibinfo {author} {\bibfnamefont {F.~M.}\ \bibnamefont {Surace}},
  \bibinfo {author} {\bibfnamefont {T.}~\bibnamefont {Schuster}}, \bibinfo
  {author} {\bibfnamefont {A.~L.}\ \bibnamefont {Shaw}}, \bibinfo {author}
  {\bibfnamefont {W.}~\bibnamefont {Gong}}, \bibinfo {author} {\bibfnamefont
  {S.}~\bibnamefont {Choi}},\ and\ \bibinfo {author} {\bibfnamefont
  {M.}~\bibnamefont {Endres}},\ }\href {https://arxiv.org/abs/2510.11679}
  {\bibinfo {title} {Observation of ballistic plasma and memory in high-energy
  gauge theory dynamics}} (\bibinfo {year} {2025}),\ \Eprint
  {https://arxiv.org/abs/2510.11679} {arXiv:2510.11679 [quant-ph]} \BibitemShut
  {NoStop}%
\bibitem [{\citenamefont {Erd{\H{o}}s}\ \emph {et~al.}(2004)\citenamefont
  {Erd{\H{o}}s}, \citenamefont {Salmhofer},\ and\ \citenamefont
  {Yau}}]{Erd_s_2004}%
  \BibitemOpen
  \bibfield  {author} {\bibinfo {author} {\bibfnamefont {L.}~\bibnamefont
  {Erd{\H{o}}s}}, \bibinfo {author} {\bibfnamefont {M.}~\bibnamefont
  {Salmhofer}},\ and\ \bibinfo {author} {\bibfnamefont {H.-T.}\ \bibnamefont
  {Yau}},\ }\bibfield  {title} {\bibinfo {title} {{On the Quantum Boltzmann
  Equation}},\ }\href {https://doi.org/10.1023/b:joss.0000037224.56191.ed}
  {\bibfield  {journal} {\bibinfo  {journal} {Journal of Statistical Physics}\
  }\textbf {\bibinfo {volume} {116}},\ \bibinfo {pages} {367} (\bibinfo {year}
  {2004})}\BibitemShut {NoStop}%
\bibitem [{\citenamefont {Berdanier}\ \emph {et~al.}(2019)\citenamefont
  {Berdanier}, \citenamefont {Scaffidi},\ and\ \citenamefont
  {Moore}}]{Berdanier_2019}%
  \BibitemOpen
  \bibfield  {author} {\bibinfo {author} {\bibfnamefont {W.}~\bibnamefont
  {Berdanier}}, \bibinfo {author} {\bibfnamefont {T.}~\bibnamefont
  {Scaffidi}},\ and\ \bibinfo {author} {\bibfnamefont {J.~E.}\ \bibnamefont
  {Moore}},\ }\bibfield  {title} {\bibinfo {title} {Energy drag in
  particle-hole symmetric systems as a quantum quench},\ }\bibfield  {journal}
  {\bibinfo  {journal} {Physical Review Letters}\ }\textbf {\bibinfo {volume}
  {123}},\ \href {https://doi.org/10.1103/physrevlett.123.246603}
  {10.1103/physrevlett.123.246603} (\bibinfo {year} {2019})\BibitemShut
  {NoStop}%
\bibitem [{\citenamefont {Young}\ \emph {et~al.}(2025)\citenamefont {Young},
  \citenamefont {Lloyd},\ and\ \citenamefont {von
  Keyserlingk}}]{young2025breakdownfermisgoldenrule}%
  \BibitemOpen
  \bibfield  {author} {\bibinfo {author} {\bibfnamefont {T.}~\bibnamefont
  {Young}}, \bibinfo {author} {\bibfnamefont {J.}~\bibnamefont {Lloyd}},\ and\
  \bibinfo {author} {\bibfnamefont {C.}~\bibnamefont {von Keyserlingk}},\
  }\href {https://arxiv.org/abs/2508.00254} {\bibinfo {title} {{Breakdown of
  Fermi's Golden Rule in 1d systems at non-zero temperature}}} (\bibinfo {year}
  {2025}),\ \Eprint {https://arxiv.org/abs/2508.00254} {arXiv:2508.00254
  [quant-ph]} \BibitemShut {NoStop}%
\bibitem [{\citenamefont {Haldane}(1981)}]{haldane1981luttinger}%
  \BibitemOpen
  \bibfield  {author} {\bibinfo {author} {\bibfnamefont {F.~D.~M.}\
  \bibnamefont {Haldane}},\ }\bibfield  {title} {\bibinfo {title} {{'Luttinger
  liquid theory'of one-dimensional quantum fluids. I. Properties of the
  Luttinger model and their extension to the general 1D interacting spinless
  Fermi gas}},\ }\href@noop {} {\bibfield  {journal} {\bibinfo  {journal}
  {Journal of Physics C: Solid State Physics}\ }\textbf {\bibinfo {volume}
  {14}},\ \bibinfo {pages} {2585} (\bibinfo {year} {1981})}\BibitemShut
  {NoStop}%
\end{thebibliography}%

\onecolumngrid
\appendix
\section{Derivation and properties of optimal models}
In this Appendix, we summarize the derivation of the optimal models Eq. \eqref{eq:trueoptimum} and discuss some of their mathematical properties as the interaction range $R \to \infty$. 

The terms in the Lagrangian Eq. \eqref{eq:Lfn} are given explicitly by Eq. \eqref{eq:exactvariance} for the objective function and
\begin{align}
\langle\hat{V}^2\rangle_{\beta=0} = \frac{L}{4} \sum_{r>0} U_r^2, \quad \langle \hat{J}^2 \rangle_{\beta=0} = \frac{L}{2} \sum_{r>0} r^2t_r^2
\end{align}
for the constraints. Solving first for stationary points with $U_r=t_r=0$ for $r>2$, we find that the only non-trivial solutions have $t_2=0$ and $|t_1|=1$. This suggests that the most effective method for attaining minimal current decay subject to the chosen constraints is through hopping that is only nearest-neighbour in combination with long-range interactions. We therefore set $t_1=t = 1$, $t_r=0$ for all $r\neq 1$, and consider Hamiltonians with non-zero density-density interactions up to some specified interaction range $1 \leq R < \lfloor L/2 \rfloor$.

In this case the Lagrangian for current decay can be written as a function of the vector of non-zero couplings $\vec{U} = (U_1, U_2,\ldots,U_R)$, as
\begin{align}
\mathcal{L}(t,\vec{U},\lambda_1,\lambda_2) = &L t^2\left(\sum_{x=1}^{R-1}(U_{x+1}-U_{x})^2+ U_R^2\right) + \frac{L\lambda_1}{4}\left(\sum_{x=1}^R U_x^2-1\right) + \frac{L\lambda_2}{2} \left(t^2-1\right).
\end{align}

Remarkably, this specific instance of the optimization problem posed above is quadratic and moreover solvable in closed form. The key observation is that the stationary condition in $\vec{U}$ can be written as a matrix eigenvalue equation $A \vec{U} = \alpha \vec{U}$, where $\alpha=- \frac{\lambda_1}{4t^2}$ and the $R$-by-$R$ matrix $A$
\begin{equation}
A = \begin{pmatrix} 1 & -1 & 0 & 0 & \ldots & 0 & 0 & 0\\
-1 & 2 & -1 & 0 &\ldots & 0 & 0 & 0 \\
0 & -1 & 2 & -1 &\ldots & 0 & 0 & 0 \\
\vdots & \vdots & \vdots & \vdots & \ddots & \vdots & \vdots & \vdots \\
0 & 0 & 0 & 0 & \ldots & -1 & 2 & -1 \\
0 & 0 & 0 & 0 & \ldots & 0 & -1 & 2 \end{pmatrix}.
\end{equation}
The eigenvalue equation fixes the value of $\lambda_1/t^2$ and the direction of the vector $\vec{U}$. The stationary condition for $\lambda_1$ imposes unit normalization of $\vec{U}$,  while the stationary conditions for $\lambda_2$ and $t$ fix $t^2 = 1$ and the numerical value of $\lambda_2$ respectively. The spectral problem for $A$ amounts to a boundary value problem for a standard tight-binding-type second-order difference equation, explicitly
\begin{align}
(1-\alpha)U_{1} - U_{2} &= 0, \\
(2-\alpha)U_{x} - U_{x-1}-U_{x+1} &= 0, \quad 1<x<R, \\
(2-\alpha)U_R - U_{R-1} &= 0.
\end{align}
We find $R$ linearly independent normalized solutions, which can be written as
\begin{equation}
U^{(m)}_x = \frac{2}{\sqrt{2R+1}}\cos{k_m(x-1/2)}, \quad 1 \leq x \leq R,
\end{equation}
with
\begin{equation}
k_m = \frac{(2m+1)\pi}{2R+1}, \quad m = 0,1,\ldots,R-1,
\end{equation}
and associated eigenvalues $\alpha_m = 2 - 2 \cos{k_m}$. The eigenvalue $\alpha_m$ is related to the decay rate of the current via $\langle \dot{\hat{J}}^2 \rangle = \alpha_m L$, and we deduce that the solution with the slowest current decay, subject to the constraints, is given by setting $m=0$ in the above expressions.

For reference, the density-density interaction strengths at range $r$ for the optimal models, $U_r^*$, are tabulated below in Table \ref{tab:Tab1} for the values $1 \leq R \leq 7$ of $R$ simulated in this paper.
\begin{table}[h]
    \centering
    \begin{tabular}{c|c|c|c|c|c|c|c|c}
    Interaction range $R$ & $U^*_1$ & $U^*_2$ & $U^*_3$ & $U^*_4$ & $U^*_5$ & $U^*_6$ & $U^*_7$ & $\langle \dot{\hat{J}}^2 \rangle$  \\
    \hline
         1 & 1 & 0 &  0 & 0 & 0 & 0 & 0 & $L$ \\
         2 & 0.851 & 0.526 &  0 & 0 & 0 & 0 & 0 & $0.382L$ \\
         3 & 0.737 & 0.591 & 0.328 & 0 & 0 & 0 & 0 &  $0.198L$ \\
         4 & 0.657 & 0.577 & 0.429 & 0.228 & 0 & 0 & 0 & $0.121L$ \\
         5 & 0.597 & 0.549 & 0.456 & 0.326 & 0.170 & 0 & 0 & $0.081L$ \\
         6 & 0.551 & 0.519 & 0.457 & 0.368 & 0.258 & 0.133 & 0 & $0.058L$ \\
         7 & 0.514 & 0.491 & 0.447 &  0.384 & 0.304 & 0.210 &
 0.107 & $0.044L$
    \end{tabular}
    \caption{Structure of optimal density-density interactions for small interaction ranges.}
    \label{tab:Tab1}
\end{table}

We next discuss the behaviour of the optimal models as $R \to \infty$ in more detail. It will be useful to let $L \to \infty$ and parameterize these models as finitely supported sequences 
\begin{equation}
s_R = (U^*_1(R),U^*_2(R),\ldots,U^*_R(R),0,0,\ldots)
\end{equation}
in the sequence space $\mathcal{S} = \mathbb{R}^{\mathbb{N}}$. Given that the effective lifetime $\tau_{\mathrm{eff}}$ of the charge current diverges in this limit, it is natural to ask whether there is any meaningful sense in which the sequences $s_R$ ``converge'' to an integrable model as $R \to \infty$.

A mathematically precise answer to this question is that while the sequences $s_R$ converge pointwise to the zero sequence $(0,0,\ldots)$ (i.e. the non-interacting limit) as $R \to \infty$, they do not converge to the zero sequence with respect to the $\ell^2$ norm $\|.\|_2$ on $\mathcal{S}$, because in this norm
\begin{equation}
\|s_R \|_2^2 = \sum_{x=1}^{\infty} U^*_x(R)^2 = 1
\end{equation}
by the constraint on the interaction strength.

In more physical terms, one can argue that the optimal models are always in a strongly interacting regime because the Fourier transform $|U(q)|^2$, where $U(q)= \sum_{x} e^{-iqx} U^*_x$, tends to a delta function of width $\sim 1/R$ about $q=0$ as $R \to \infty$, see Fig.~\ref{FigSqrtDelta} and the discussion below. Since $|U(q)|^2$ sets the rate of scattering by $\hat{V}$, that appears for example in the collision integral of the quantum Boltzmann equation\cite{Erd_s_2004}, the optimal models always exhibit strong particle-particle scattering. Note that this is consistent with what is observed in Fig. \ref{Fig2} of the main text, and remains true even when the pseudomomentum transferred by these individual scattering events is very small.

Thus in some physically important aspects, these optimal models do not converge to their non-interacting limit as $R \to \infty$.  It seems to us that the most natural expectation is that the optimal models remain chaotic for $R \gg 1$. 

\section{Quantum Boltzmann equation for hot band sound}
In this Appendix, we argue that the dynamics of optimal models for $R \gg 1$ and infinite temperature is approximated by a linearized Boltzmann equation of the form
\begin{equation}
\label{eq:effhydro}
\partial_t \delta \rho_k +v_k \partial_x \delta \rho_k =  \partial_k\left(\frac{D}{|\cos{k}|}\delta \rho_k\right),
\end{equation}
where the effective pseudomomentum diffusion constant $D \propto \frac{1}{R}$. Note that since we are in a band, $k$-space is finite and periodic. This equation locally conserves the distribution function $\delta \rho_k(x,t)$ in phase space, and also conserves fluctuations in the local particle density $\delta n(x,t) = \int_{-\pi}^{\pi} \frac{dk}{2\pi}\, \delta \rho_k(x,t)$ and in the local current density $\delta j(x,t) = \int_{-\pi}^{\pi} \frac{dk}{2\pi} \, \delta \rho_k(x,t) v_k$, which implies that it supports ballistic sound modes. Eq. \eqref{eq:effhydro} further predicts a $k$-space diffusion time $\tau \propto R$ at which Umklapp processes become important. As we shall show below, this equation pertains only to modes that are even in a reflection of the Brillouin zone about $k=\pi/2$. We find that modes that are odd under such a reflection, including the energy density, do not relax \emph{at all} at the level of approximation of Boltzmann theory; we leave a more detailed exploration of this non-trivial physics to future work.

To derive Eq. \eqref{eq:effhydro}, let us consider an optimal model on $L$ sites with interactions of range $ 1 \ll R < \lfloor L/2 \rfloor$, as defined in the main text. Discarding terms that commute with the Hamiltonian, the interaction term can be written in terms of Fourier modes as
\begin{equation}
\hat{V} = \frac{1}{L}\sum_{q,k,k'} U(q) \hat{c}_{k+q}^\dagger \hat{c}_{k'-q}^\dagger \hat{c}_{k'} \hat{c}_{k}
\end{equation}
where a direct calculation reveals that
\begin{align}
\nonumber U(q) = \sum_{x}e^{-iqx}U^*_x = \frac{2}{\sqrt{2R+1}}\Bigg[& \frac{\sin{\left( \frac{\pi}{2R+1}+q\right)\left(\frac{R}{2}\right)}}{\sin{\left(\frac{\pi}{2R+1}+q\right)\left(\frac{1}{2}\right)}} \cos{\left(\frac{\pi R}{2(2R+1)}+q\left(\frac{R+1}{2}\right) \right)} \\
- &\frac{\sin{\left( \frac{\pi}{2R+1}-q\right)\left(\frac{R}{2}\right)}}{\sin{\left( \frac{\pi}{2R+1}-q\right)\left(\frac{1}{2}\right)}} \cos{\left( \frac{\pi R}{2(2R+1)}-q\left(\frac{R+1}{2}\right)\right)}\Bigg].
\end{align}

\begin{figure}[t]
    \centering
\includegraphics[width=0.45\linewidth]{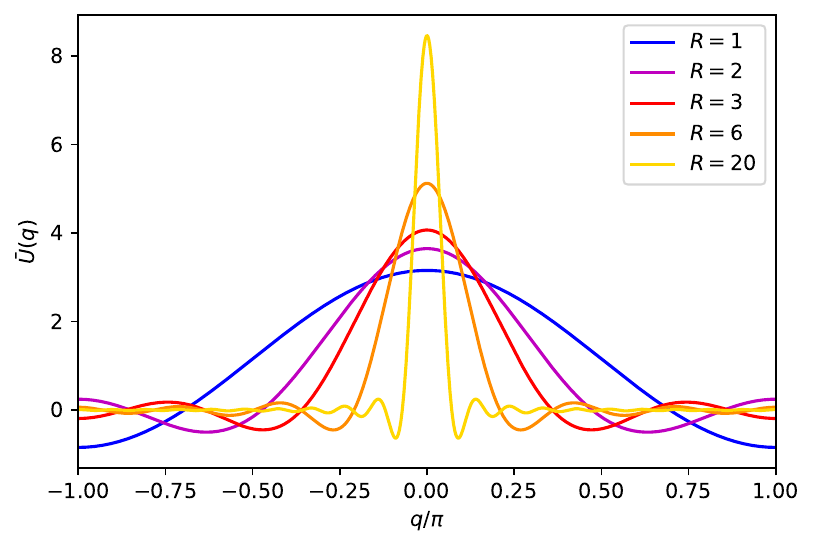}    \includegraphics[width=0.45\linewidth]{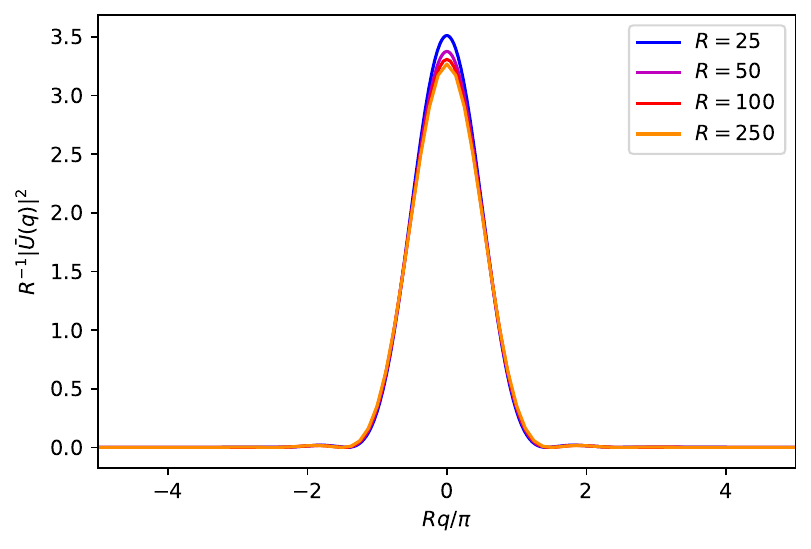}
    \caption{\emph{Left:} Fourier transform of the interaction potential for optimal models with various ranges $R$. It is clear that the Fourier transform becomes increasingly localized about $q=0$ with increasing $R$. \emph{Right:} scaling collapse of $|\bar{U}(q)|^2$ to a delta function as $R \to \infty$. The scaling of the axes demonstrates that $|\bar{U}(q)|^2$ has height $\propto R$ and support $\propto 1/R$ as $R \to \infty$. The area under these curves is independent of $R$ by Parseval's theorem and our real-space constraint $\sum_{n\geq 1}|U_n|^2 = 1$ on the interactions.}
    \label{FigSqrtDelta}
\end{figure}

In order to obtain a continuous long-wavelength limit of $U_x^*(R)$, we regulate its discontinuity at $x=0$ by introducing the function
\begin{equation}
\bar{U}_x(R) = U^*_x(R) + \frac{2}{\sqrt{2R+1}}\delta_{x0}
\end{equation}
(this range-zero interaction term yields a multiple of the number operator and does not change the dynamics). This implies that the long-wavelength $(L,R \gg 1)$ behaviour due to interactions is controlled by the function
\begin{equation}
\bar{U}(q) = U(q) + \frac{2}{\sqrt{2R+1}}.
\end{equation}
Note that $\bar{U}(q)$ becomes increasingly localized about $q=0$ as $R$ is increased (intuitively speaking, $\bar{U}(q)$ tends to the ``square root'' of a Dirac delta function as $R \to \infty$). 

In the thermodynamic limit $L \to \infty$, a microscopic derivation of the quantum Boltzmann equation~\cite{Erd_s_2004} yields ($\hbar=1$)
\begin{equation}
\partial_t \rho_k + v_k \partial_x \rho_k = 4 \pi \int \frac{dk_2dk_3dk_4}{(2\pi)^2}|V(k,k_2,k_3,k_4)|^2 \delta(k+k_2-k_3-k_4) \delta(\varepsilon_k+\varepsilon_{k_2}-\varepsilon_{k_3}-\varepsilon_{k_4})\left(\rho_{k_4}\rho_{k_3}\bar{\rho}_{k_2}\bar{\rho}_k -  \rho_k \rho_{k_2} \bar{\rho}_{k_3}\bar{\rho}_{k_4}\right),
\end{equation}
where the single-particle dispersion $\varepsilon_k = -2\cos{k}$ in this case, $\bar{\rho}_k := 1-\rho_k$, the delta-function constraint on momenta has the periodicity of the reciprocal lattice, the integrals are over the single-particle Brillouin zone $\mathrm{B.Z.}=[-\pi,\pi)$, and $|V(k_1,k_2,k_3,k_4)|^2 = |\bar{U}(k_1-k_4)-\bar{U}(k_2-k_4)|^2$ is the appropriately antisymmetrized matrix element for fermionic interactions. Using the momentum-space delta function to eliminate $k_3 = k+k'-k_4+Q$, where the requirement that $k_3 \in \mathrm{B.Z.}$ picks out a unique reciprocal lattice vector $Q \in 2 \pi \mathbb{Z}$, we deduce that
\begin{align}
\nonumber 
\partial_t \rho_k + v_k \partial_x \rho_k = \int \frac{dk' dk_4}{\pi}&|\bar{U}(k-k_4)-\bar{U}(k'-k_4)|^2 \delta(\varepsilon_{k}+\varepsilon_{k'}-\varepsilon_{k+k'-k_4}-\varepsilon_{k_4})\\
&\left(\rho_{k_4}\rho_{k+k'-k_4+Q}\bar{\rho}_{k'}\bar{\rho}_k -  \rho_k \rho_{k'} \bar{\rho}_{k+k'-k_4+Q}\bar{\rho}_{k_4}\right).
\end{align}
It is convenient to write the energy-conservation constraint in terms of the variable $q=k_4-k$ and as a constraint on $k'$, which yields
\begin{equation}
\delta(\varepsilon_{k+q}+\varepsilon_{k'-q}-\varepsilon_{k'}-\varepsilon_{k}) =\frac{1}{2|\sin{(k+q)}-\sin{k}|}\left(\delta(k'-k-q)+\delta(k'-\pi+k)\right)
\end{equation}
where the delta functions again have the periodicity of the reciprocal lattice. The first term yields a vanishing contribution to the collision integral, since elastic momentum exchange cannot produce entropy. The second term is non-vanishing and yields
\begin{equation}
\label{eq:effectiveKE}
\partial_t \rho_k + v_k \partial_x \rho_k = \int_{-\pi}^{\pi} \frac{dk_4}{2\pi} \, \frac{|\bar{U}(k-k_4)-\bar{U}(k^*-k_4)|^2}{|\sin{k}-\sin{k_4}|} \left(\rho_{k_4}\rho_{k_4^*}\bar{\rho}_{k^*}\bar{\rho}_k -  \rho_k \rho_{k^*} \bar{\rho}_{k_4^*}\bar{\rho}_{k_4}\right),
\end{equation}
where in this expression, $k^* = \pi-k+Q'$ denotes the unique $k^* \in \mathrm{B.Z.}$, $k^*\neq k$, such that $v_k=v_{k^*}$, where the reciprocal lattice vector $Q' = 0$ for $k >0$ and $Q'=-2\pi$ otherwise. Thus the dominant contribution to relaxation at short times and pseudomomentum $k$ is the scattering of pairs of quasiparticles moving at the same group velocity. Precisely these processes were recently identified as being responsible for the singular corrections to Fermi's Golden Rule in one spatial dimension~\cite{Berdanier_2019,young2025breakdownfermisgoldenrule}.

To proceed further, note that the ``conjugate'' momentum $k^*$ satisfies
\begin{equation}
\partial_t \rho_{k^*} + v_k \partial_x \rho_{k^*} = I_{k}[\rho],
\end{equation}
where $I_k[\rho]$ denotes the right-hand side of Eq. \eqref{eq:effectiveKE}. Defining even and odd (about reflection in $k=\pi/2$) single-particle distribution functions $\rho^{\pm}_k = \frac{1}{2}(\rho_k \pm \rho_{k^*})$ leads us to the remarkable conclusion that at this level of approximation, the odd parts of the single-particle distribution function do not relax,
\begin{equation}
\partial_t \rho^-_k + v_k \partial_x \rho^-_k = 0
\end{equation}
\emph{regardless} of the choice of interaction potential, and instead propagate ballistically. Thus the dissipative relaxation of these modes in one dimension is an effect beyond conventional Boltzmann theory, and presumably requires keeping track of higher-order terms in the interaction strength (i.e. going beyond the second Born approximation). 

\begin{figure}[t]
    \centering \includegraphics[width=0.32\linewidth]{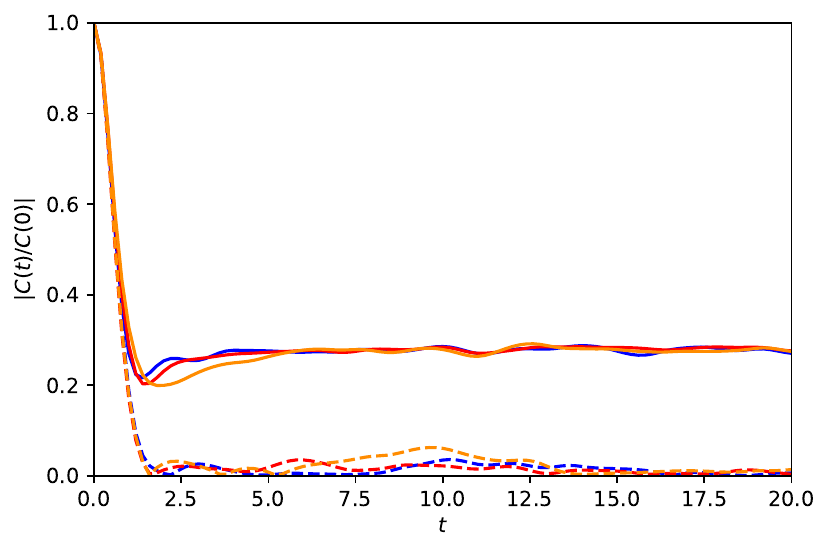}
\includegraphics[width=0.32\linewidth]{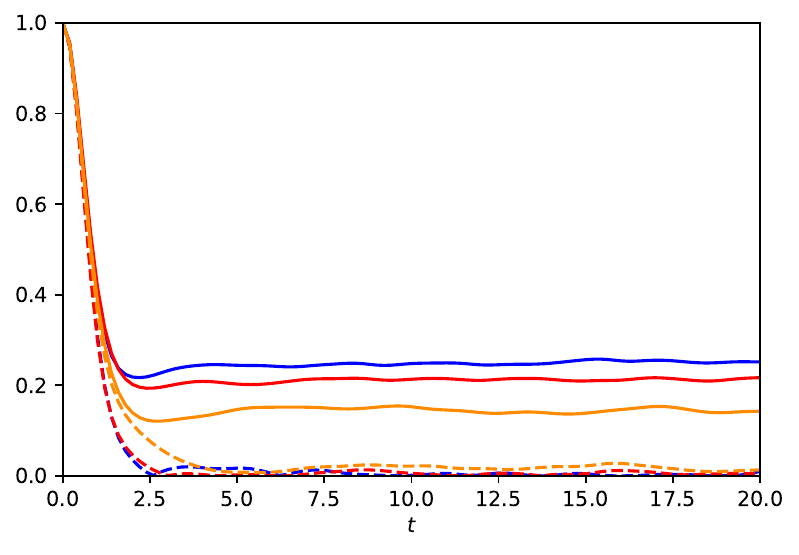}
\includegraphics[width=0.32\linewidth]{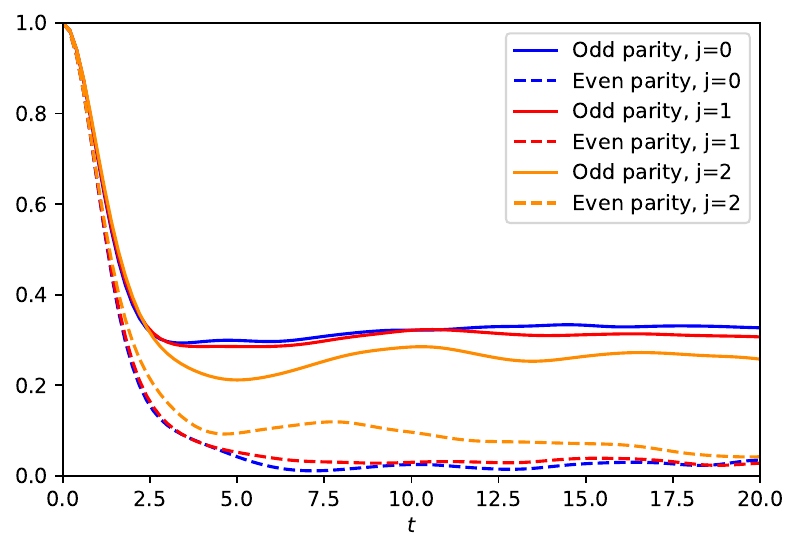}
    \caption{Visibly different relaxation between even and odd parity modes with respect to reflection in $k=\pi/2$. We simulate optimal models on $L=14$ sites at half filling with interaction ranges $R=1$ (\emph{left}) corresponding to the integrable case, $R=3$ (\emph{middle}) and $R=6$ (\emph{right}) and let $k=2\pi j/L$ for $j=0,1,2$. We consider infinite-temperature two-time correlation functions $C(t) = \langle \hat{O}(t) \hat{O}(0)\rangle_{\beta=0}$ of traceless even-parity operators $\hat{O}=  \hat{n}_k+\hat{n}_{k^*} - \mathbbm{1}$ and traceless odd-parity operators $\hat{O} = \hat{n}_k - \hat{n}_{k^*}$ at the $k$ points in question. Discrepancies in relaxation rates between odd and even modes for all $R$, together with increasing relaxation times of even-parity modes with increasing $R$, are apparent. Both properties are non-trivial predictions of the linearized kinetic theory Eqs.~\eqref{eq:linKEodd} and ~\eqref{eq:linKEeven}.}
    \label{fig:oddeven}
\end{figure}

Meanwhile the even parts
\begin{equation}
\partial_t \rho^+_k + v_k \partial_x \rho^+_k = I_k[\rho] 
\end{equation}
relax on a generically faster timescale set by normal scattering processes. (This ceases to hold in the vicinity of the points $k=\pm \pi/2$, for which $k^*=k$ and the Boltzmann collision integral vanishes, just as for the odd-parity modes.) This is reminiscent of the scenario of ``tomographic dynamics'' in 2D Fermi liquids, whereby the constraints arising from Fermi-Dirac statistics at low temperature imply that odd-parity excitations of the Fermi surface relax much slower than their even-parity counterparts~\cite{Ledwith_2019}. However, in our case there is no low-temperature constraint and parity-dependent relaxation follows from the kinematic constraints imposed by energy and momentum conservation in one dimension. This discrepancy between the relaxation rates of odd and even parity modes is apparent even upon exact diagonalization of small systems, see Fig. \ref{fig:oddeven}.

In order to compare this understanding to the results of the main text, and given the temperature-independence of these results, we now linearize about the half-filled, infinite-temperature state $\rho_k = \frac{1}{2}+\delta\rho_k$. It follows by the above considerations that the linearized odd-parity modes satisfy the free-streaming Boltzmann equation
\begin{equation}
\label{eq:linKEodd}
\partial_t \delta\rho_k^- + v_k \partial_x \delta\rho_k^- = 0,
\end{equation}
while the linearized even-parity modes satisfy
\begin{equation}
    \partial_t \delta \rho^+_k + v_k\partial_x \delta \rho^+_k = \int_{-\pi}^{\pi} \frac{dk_4}{4\pi} \, \frac{|U(k-k_4)-U(k^*-k_4)|^2}{|\sin{k_4}-\sin{k}|} (\delta \rho^+_{k_4}-\delta \rho^+_k).
\end{equation}
Away from the singular points $k = \pm\pi/2$, at which $\delta \rho_{k}^+$ is quasiconserved, the dominant contributions to the linearized collision integral are from regions centered on $k_4 = k$ and $k_4 = k^*$ with characteristic widths $\sim 1/R$. Expanding to leading non-trivial order in $q$, we find that
\begin{equation}
\partial_t \delta \rho^+_k + v_k\partial_x \delta \rho^+_k \approx  \int_{-\pi}^{\pi} \frac{dq}{4\pi} \frac{|q||U(q)|^2}{|\cos{k}|}(\tan{k}\,\partial_k \delta \rho_k^+ + \partial_k^2 \delta \rho_k^+),
\end{equation}
where discarding terms of $\mathcal{O}(q^2)$ under the integral sign is justified by the scaling
\begin{equation}
\label{eq:scalinginR}
\int_{-\pi}^{\pi} dq \, |q|^n |U(q)|^2 \sim R^{-n} \int_{-\infty}^{\infty} dq' \, {|q'|}^nf(q') \propto R^{-n}, \quad R \gg 1,
\end{equation}
which follows by the emergence of a large-$R$ scaling function $f(q') = \lim_{R\to\infty} R^{-1}|\bar{U}(q'/R)|^2$ as in Fig. \ref{FigSqrtDelta}.

Defining the constant $D = \int_{-\pi}^{\pi} \frac{dq}{4\pi}{|q||U(q)|^2}$, this yields the inhomogeneous diffusion equation
\begin{equation}
\label{eq:linKEeven}
\partial_t \delta \rho_k^+ + v_k \partial_x \delta \rho_k^+ =  \partial_k \left(D_k\partial_k \delta \rho_k^+\right), \quad D_k = \frac{D}{|\cos{k}|},
\end{equation}
away from the singular points, i.e. in the bulk of the Brillouin zone $|k \pm \pi/2| \gg 1/R$. The linearized kinetic equations \eqref{eq:linKEodd} and \eqref{eq:linKEeven}, which might have been difficult to guess \emph{a priori}, have several appealing and non-trivial properties.

Most importantly, they give rise to local conservation laws for charge and for a momentum-like degree of freedom, as required for a physical interpretation in terms of an emergent sound mode. To see this, consider fluctuations
$\delta n(x,t) = \int_{-\pi}^{\pi} \frac{dk}{2\pi} \, \delta \rho_k(x,t)$, $\delta j(x,t) =  \int_{-\pi}^{\pi} \frac{dk}{2\pi} \, \delta \rho_{k}(x,t)v_k$ and $\delta e(x,t) = \int_{-\pi}^{\pi} \frac{dk}{2\pi} \, \delta \rho_{k}(x,t)e_k$ in the local charge, charge current and energy density respectively. Charge and energy are conservation laws of the microscopic dynamics, while charge current is not, but turns out to be an emergent conservation law of Eq. \eqref{eq:linKEeven}. First, note that because density and velocity are even with respect to inversion about $k=\pi/2$ and energy is odd, it follows that fluctuations of the energy density are quasiconserved and decoupled from $\delta n$ and $\delta j$ (thus our kinetic model additionally predicts quasiballistic energy transport). Meanwhile, fluctuations of particle density and velocity density are locally conserved because the resulting collision integrands are total derivatives away from the singular points in the Brillouin zone (since the full collision integral in Eq. \eqref{eq:effectiveKE} vanishes at the singular points, their contributions can safely be ignored). Thus our kinetic theory predicts that the optimal models indeed support coupled ballistic hydrodynamics of charge and charge current, which is the physics of sound.
Moreover, these are the only hydrodynamic conservation laws consistent with Eq. \eqref{eq:linKEeven} in the even-parity sector, since any locally conserved even-parity charge density $\delta q(x,t) = \int_{-\pi}^{\pi} \frac{dk}{2\pi}\delta\rho^+_k(x,t)q_k$ must satisfy $q_k' \propto \cos{k}$ for $|k| \neq \pi/2$.

We will largely leave a more detailed analysis of this kinetic theory to future work, except to note that the fastest relaxing modes resulting from Eqs. \eqref{eq:linKEodd} and \eqref{eq:linKEeven} are the $k=0,\pm\pi$ modes with $D_k = D$. Thus the characteristic timescale for diffusion in momentum space is set by $D$. We further note that by Eq. \eqref{eq:scalinginR}, $D \propto 1/R$ for $R \gg 1$, implying scaling $\tau \propto 1/R$ of the relaxation time that is consistent with our very different method of analysis in the main text.

\section{Numerical results away from infinite temperature}

\begin{figure}[t]
    \centering
\includegraphics[width=0.49\linewidth]{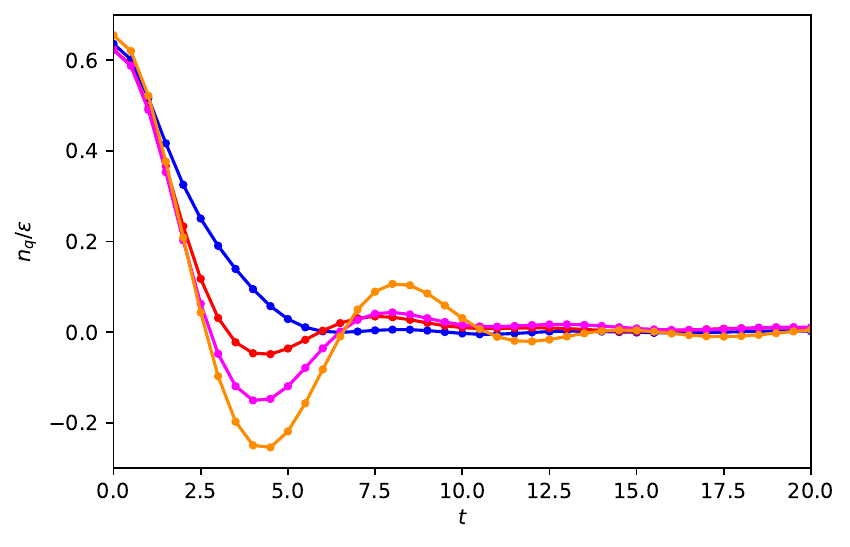}
\includegraphics[width=0.49\linewidth]{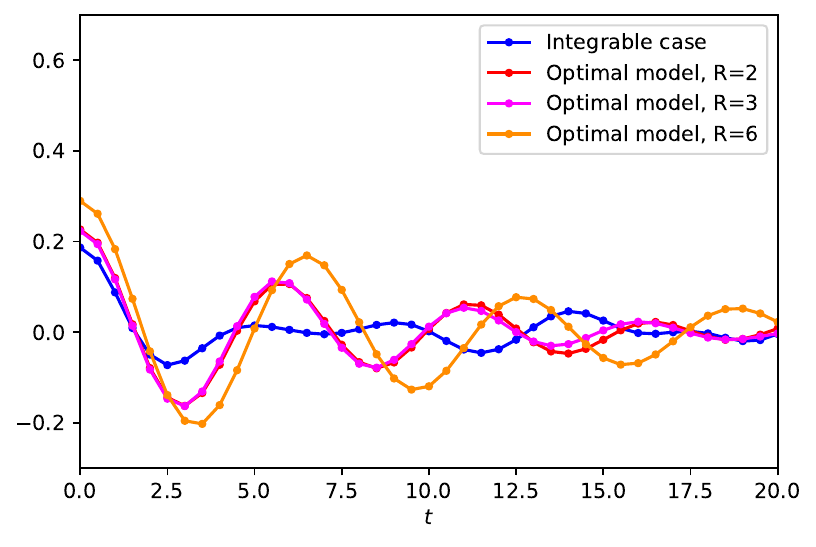}
    \caption{Decay of an initial density modulation in a system of $L=14$ sites at half filling and inverse temperature $\beta=0.1$ (\emph{left}) and $\beta=1$ (\emph{right}), for the same model parameters as Fig. \ref{Fig1} of the main text and the initial condition Eq. \eqref{eq:initialdensmodT} with $\epsilon = 0.01$ and $q = 2\pi/L$. It is clear from this plot that the phenomenology of hot band sound survives at finite temperature.}
    \label{fig:finiteT}
\end{figure}

In this Appendix, we show that the effect identified in the main text persists away from infinite temperature, which is consistent with the generic physical mechanism of slow diffusion in pseudomomentum space identified above. Specifically, we simulate a family of states that generalize Eq. \eqref{eq:initialdensmod} to finite temperatures, namely
\begin{equation}
\label{eq:initialdensmodT}
\hat{\rho}(0) = \frac{1}{Z}\exp\left(-\beta \hat{H} + \epsilon \sum_{x=1}^{L} \sin{(qx)} (\hat{n}_x - \langle \hat{n}_x \rangle_{\beta})\right),
\end{equation}
with $Z = \mathrm{tr}[\hat{\rho}(0)]$ and parameters $\epsilon=0.01$, $q = 2\pi/L$ and $L=14$ as in Fig. \ref{Fig1}, except that we now set $\beta = 0.1$ and $\beta = 1$. We note that these states can equivalently be viewed as resulting from weakly inhomogeneous chemical potentials $\mu(x) = \frac{\epsilon}{\beta} \sin{qx}$ at inverse temperature $\beta$. It is clear from Fig.~\ref{fig:finiteT} that underdamped oscillations of an initial charge density modulation persist (and are indeed enhanced) at such high and intermediate temperatures on the scale of the bandwidth. We do not consider low temperatures, where one-dimensional metals exhibit emergent integrability in the form of Luttinger liquid physics~\cite{haldane1981luttinger} and such underdamped relaxation would not be surprising.

\end{document}